\documentclass[12pt,letterpaper]{article}

\usepackage{amsmath}
\usepackage{amssymb}
\usepackage{epsfig}
\usepackage{graphicx}
\usepackage{cite}

\newcommand{\capdef}{}
\newcommand{\mycaption}[2][\capdef]{\renewcommand{\capdef}{#2}%
       \caption[#1]{{\footnotesize #2}}}
\makeatletter
\renewcommand{\fnum@table}{\textbf{\tablename~\thetable}}
\renewcommand{\fnum@figure}{\textbf{\figurename~\thefigure}}
\makeatother

\newcounter{myenumi}

\renewcommand{\themyenumi}{\roman{myenumi}}
{\end{list}}

\newlength{\myem}
\newcounter{mysubequation}[equation]

\makeatletter
\renewcommand{\section}{\@startsection{section}{1}{0em}{-\baselineskip}%
{\baselineskip}{\normalfont\large\bfseries}}
\renewcommand{\subsection}%
{\@startsection{subsection}{2}{0em}{-0.7\baselineskip}%
{0.7\baselineskip}{\normalfont\bfseries}}
\makeatother

\newcommand{\ie}{{\it i.e.}\ }

\newcommand{\eg}{{\it e.g.}\ }

\newcommand{\eq}{Eq.}
\newcommand{\eqs}{Eqs.}

\newcommand{\fig}{Figure}

\newcommand{\Refs}{Refs.}

\newcommand{\KtoK}{\mbox{\sf K2K}}
\newcommand{\JHFSK}{\mbox{\sf T2K}}

\newcommand{\MINOS}{\mbox{\sf MINOS}}

\newcommand{\SK}{\mbox{\sf Super-K}}

\newcommand{\globes}{\mbox{\sf GLoBES}\ }
\newcommand{\stheta}{\sin^22\theta_{13}}
\newcommand{\deltacp}{\delta_\mathrm{CP}}
\newcommand{\ldm}{\Delta m_{31}^2}

\newcommand{\equ}[1]{\eq~(\ref{equ:#1})}
\newcommand{\figu}[1]{\fig~\ref{fig:#1}}

\begin{document}

\begin{titlepage}

\renewcommand{\thefootnote}{\alph{footnote}}

\vspace*{-3.cm}
\begin{flushright}
STUPP-06-188\\
TUM-HEP-654/06\\

\end{flushright}

\vspace*{0.5cm}

\renewcommand{\thefootnote}{\fnsymbol{footnote}}
\setcounter{footnote}{-1}

{\begin{center}
{\Large\bf Neutrino Beams From Electron Capture\\ at High Gamma}
\end{center}}

\renewcommand{\thefootnote}{\alph{footnote}}

\vspace*{.8cm}
{\begin{center} {\large{\sc
                Mark~Rolinec\footnote[1]{\makebox[1.cm]{Email:}
                rolinec@ph.tum.de}~and~
                Joe~Sato\footnote[2]{\makebox[1.cm]{Email:}
                joe@phy.saitama-u.ac.jp}
                }}
\end{center}}
\vspace*{0cm}
{\it
\begin{center}

\footnotemark[1]${}^,$\footnotemark[3]%
       Physik--Department, Technische Universit\"at M\"unchen, \\
       James--Franck--Strasse, 85748 Garching, Germany

\vspace*{1mm}

\footnotemark[2]
       Department of Physics, Saitama University, \\
       Shimo-okubo, Sakura-ku, Saitama, 338-8570, Japan

\vspace*{1cm}

\today
\end{center}}

\vspace*{0.3cm}

\begin{abstract}
We investigate the potential of a flavor pure high gamma electron capture 
electron neutrino beam directed towards a large Water Cherenkov detector 
with 500~kt fiducial mass. The energy of the neutrinos is reconstructed 
by the position measurement within the detector and superb energy resolution 
capabilities could be achieved. We estimate the requirements for such a 
scenario to be competitive to a neutrino/anti-neutrino running at a neutrino 
factory with less accurate energy resolution. Although the requirements 
turn out to be extreme, in principle such a scenario could achieve as good 
abilities to resolve correlations and degeneracies in the search for
$\sin^22\theta_{13}$ and $\delta_{CP}$ as a standard neutrino factory
experiment.

\end{abstract}

\vspace*{.5cm}

\end{titlepage}

\newpage

\renewcommand{\thefootnote}{\arabic{footnote}}
\setcounter{footnote}{0}


\section{Introduction}
All observations on neutrinos coming from the sun\cite{Cleveland:1998nv,
Abdurashitov:2002nt,Hampel:1998xg,Altmann:2000ft, Fukuda:2001nk,Fukuda:2001nj,
Fukuda:2002pe,Ahmad:2002ka,Ahmad:2002jz,Ahmed:2003kj}, the atmosphere\cite{Fukuda:1998mi,
Fukuda:1998ah,Fukuda:2000np,Ashie:2005ik,Ambrosio:1998wu,Ronga:2001zw,
Sanchez:2003rb,Ambrosio:2004ig}, and reactors\cite{Eguchi:2002dm,Araki:2004mb} are
well understood in the picture of neutrino oscillations\cite{Maki:1962mu} in the
three generation framework of lepton mixing. Two of the mixing angles, $\sin^22\theta_{12}$ 
and $\sin^22\theta_{23}$ have been measured as well as the two mass square 
differences $|\ldm|$ and $\Delta m^2_{21}$ have been determined. 
Furthermore, the parameters which are mainly relevant in the atmospheric neutrino 
oscillations,~\ie $\sin^22\theta_{23}$ and $|\ldm|$ have been confirmed 
by the terrestrial experiments \KtoK\cite{Ahn:2002up,Oyama:2005mi,Ahn:2006zz} and 
\MINOS\cite{Tagg:2006sx,Michael:2006rx}.

However, the remaining two mixing parameters, the third mixing angle $\stheta$ and 
the CP violating phase $\deltacp$ have not been determined yet. Currently, there 
only exists an upper bound for $\stheta$\cite{Apollonio:1999ae,Apollonio:2002gd} and
there is no information on the value of $\deltacp$. Also, the sign of the mass 
squared difference $\ldm$ is currently unknown,~\ie it is unclear if neutrinos exist 
in normal or inverted hierarchy. So, the aim of future oscillation experiments is to 
measure these two parameters, to improve the precision to the leading solar and 
atmospheric parameters, and determine the neutrino mass hierarchy. In order to complete 
the picture of neutrino oscillation parameters, several types of new experiments have 
been proposed and are studied extensively. This includes reactor 
experiments\cite{Minakata:2002jv,Huber:2003pm,Anderson:2004pk,Ardellier:2004ui,
Ardellier:2006mn,Huber:2006vr} that are only sensitive to $\stheta$, and experiments 
where information on both, $\stheta$ and $\deltacp$ can be obtained, like superbeam 
experiments\cite{Minakata:2000ee,Sato:2000wv,Richter:2000pu,Ables:1995wq,Itow:2001ee,
Ambats:2004js}, neutrino factories\cite{Geer:1998iz,DeRujula:1998hd,Dick:1999ed,Barger:1999fs,Cervera:2000kp,Albright:2000xi,Blondel:2000gj,
Apollonio:2002en,Huber:2002mx}, and beta-beams\cite{Zucchelli:2002sa,Mezzetto:2003ub,
Bouchez:2003fy,Mezzetto:2004gs, Burguet-Castell:2003vv,Burguet-Castell:2005pa,
Terranova:2004hu,Donini:2004hu,Donini:2004iv,Donini:2006dx,Donini:2006tt,Huber:2005jk,
Campagne:2006yx,Volpe:2006in}.

Recently, another idea has been proposed, which makes use of a neutrino beam with neutrinos 
coming from electron capture processes~\cite{Sato:2005ma,Bernabeu:2005jh}. The electron 
neutrinos that are emitted from such electron capture processes would have a definite 
energy $Q$ in the rest frame of the mother nuclei. Therefore by accelerating the mother 
nuclei to a Lorentz factor $\gamma$ the neutrino energy $\mathrm{E_\nu}$ can be completely 
controlled, since the energy of the neutrinos that are boosted exactly towards the direction 
of the detector is $\mathrm{E_\nu}=2\gamma Q$. So, the $\gamma$ factor and the baseline 
length $L$ have to be chosen respectively to the $Q$ value of the electron capture process, 
the location of the oscillation maximum, and the minimal energy observable at the detector,~\eg 
above the Cherenkov threshold of muons at a Water Cherenkov detector. For example, if $Q$ is 
relatively large ($\mathcal{O}$(1 MeV)), $\gamma$ can be chosen to be of the order 
$\mathcal{O}$(100). In this case the neutrino beam can be viewed as exactly monoenergetic 
in the detector\cite{Sato:2005ma,Sato:2006ha,Bernabeu:2005jh,Lindroos:2005jz,Bernabeu:2005ks,
Bernabeu:2005kq,Bernabeu:2005zs,Volpe:2006in}. On the contrary, if $Q$ is relatively small 
($\mathcal{O}$(100 keV)) the $\gamma$ must be chosen quite high ($\mathcal{O}$(1000)), 
but the necessary choice of the baseline leads to the effect that the neutrinos now have 
a wider energy range at the detector. While the maximal energy of $\mathrm{E_\nu}=2\gamma Q$ 
is reached by the neutrinos in the beam axis, the energy of the neutrinos becomes smaller 
off the axis and the minimal observable energy of the neutrinos depends on the detector size and 
the baseline. In this scenario, the neutrino energy can be reconstructed from the vertex position 
measurement relatively to the beam axis within the detector and in principle a superb energy
resolution can be achieved\cite{Sato:2005ma,Sato:2006ha}. This however requires in addition to the 
resolution of the position measurement within the detector, that the beam divergence of the stored mother nuclei
can be accurately controlled. This scenario seems interesting 
since only with one acceleration factor $\gamma$ a wide range of neutrino energy can be 
covered simultaneously with a very accurate neutrino energy determination.

In this work we investigate the potential of such scenarios with a flavor pure 
electron neutrino beam coming from beta capture at high $\gamma$ lead towards a 
Water Cherenkov detector with a fiducial mass of 500~kt. We will refer to these 
scenarios as monobeam scenarios in the following. We estimate the requirements 
for such a scenario to be able to resolve correlations and degeneracies in the 
search for the remaining oscillation parameters $\stheta$ and $\deltacp$
within the measurement in only one polarity,~\ie neutrino running, but with superb 
energy resolution abilities and to be competitive to a standard neutrino factory scenario with 
neutrino and anti-neutrino running, but less accurate energy reconstruction. 
Unfortunately, the ability to also gain information on the sign of $\ldm$ at the 
discussed monobeam scenarios is limited due to the missing anti-neutrino running, 
so it will be omitted throughout this work.

This work is organized as follows: In Section 2 we sketch the basic principles of 
the high gamma electron capture monobeam experiments and summarize all underlying 
assumptions. Furthermore, we define the reference setups that are investigated 
throughout the rest of the work. Next, in Section 3 we address the issue of 
requirements to resolve parameter correlations and degeneracies in the sensitivity 
to $\stheta$ at the reference scenarios defined in Section 2 and then address the
sensitivity to $\deltacp$ in Section 4. Here, also all parameter correlations and
degeneracies are taken into account. We summarize and conclude the main results in Section 5. 
Finally, the details of the operation of a monobeam experiment including the energy 
reconstruction by the position measurement,~\ie a derivation of the neutrino energy 
$E_\nu(R)$ as a function of the radius from the beam axis, and the details 
of the event rate calculation is presented in the Appendix.  

\section{Experiment configurations and simulation techniques}

\begin{figure}[t]
\begin{center}
\includegraphics[width=10cm]{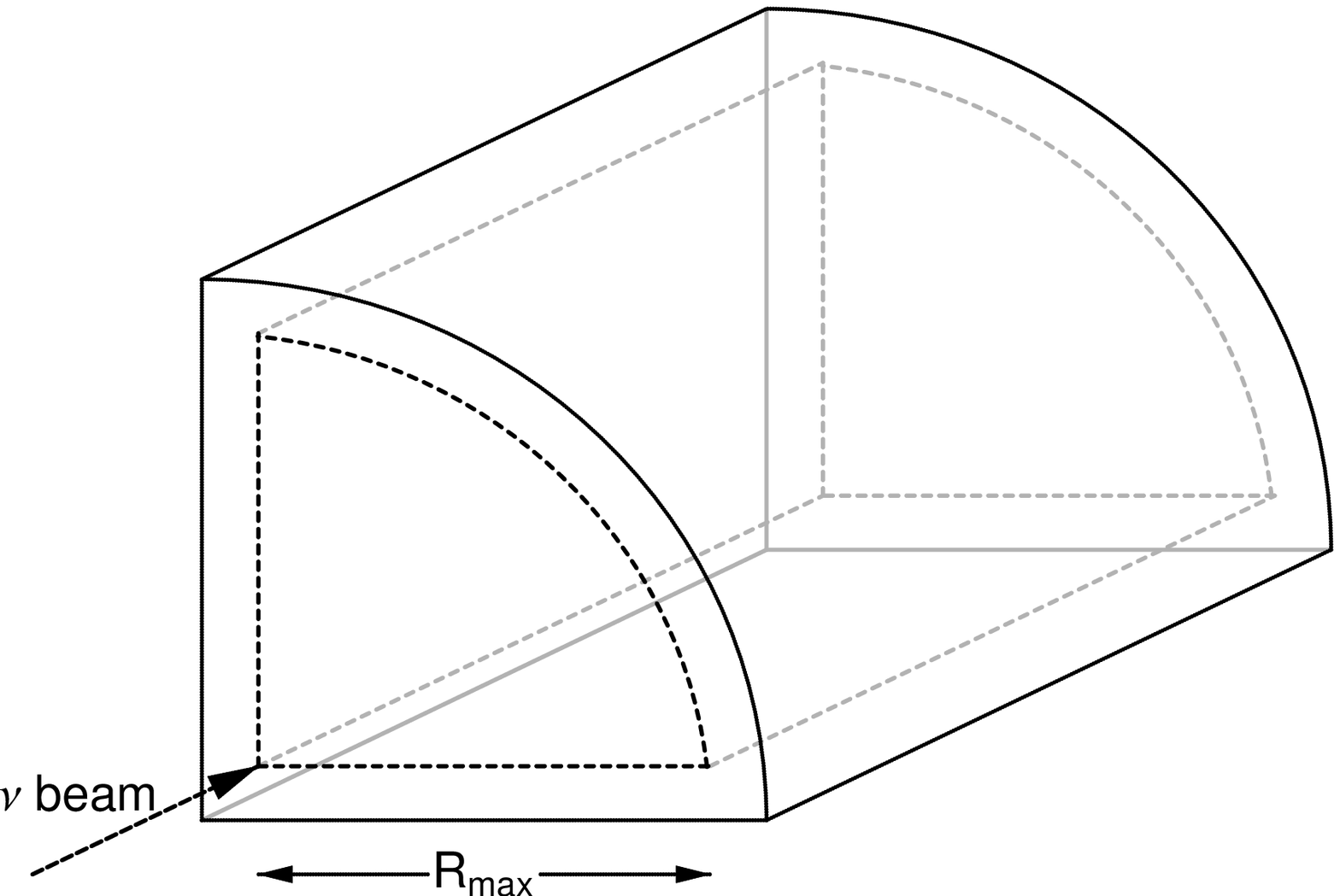}
\end{center}
\mycaption{\label{fig:detector} Scheme of the detector setting. The
fiducial volume is indicated by the dashed lines. The neutrino beam hits
the detector at the edge of the fiducial volume to allow for distance
measurements from the beam axis. In case of $R_{\mathrm{max}}=100\,\mathrm{m}$ the
depth of the fiducial volume part would be approximately 64~m for a 500~kt
fiducial detector mass.}
\end{figure}

In the scope of this work we consider a flavor pure neutrino beam that is produced 
within the electron capture process of $^{110}_{50}\mathrm{Sn}$ isotopes:
\begin{eqnarray}
\mathrm{ ^{110}_{50}Sn\,+\,e^- \,\rightarrow\,^{110}_{49}In \, + \,
\nu_e. }
\end{eqnarray}
In the rest frame of the process the produced neutrinos are monochromatic with 
an energy of Q~=~267~keV\footnote{We only consider electron capture from the K 
shell here. A more detailed analysis should also include electron capture from 
higher shells. However, the results should not be affected significantly. On one 
hand a position measurement of a neutrino would allow different true energy values 
and a new discrete uncertainty arises, but on the other hand the ratio is known 
and the higher the shell, the smaller the contribution. Furthermore, the sets of 
neutrinos from electron captures from other shells cannot be interpreted as 
background since also their energy is accurately known, besides a discrete 
uncertainty, and they also oscillate and hence contribute to the fit.} at a lifetime 
of 4.11~h\footnote{This is the main limiting factor for obtaining an adequate number of
electron capture processes per year,~\ie to collect enough statistics. However, in 
\cite{Ikeda:2005yh,Yoshimura:2005zq,Nomura:2006ik} the possibility to enhance the 
electron capture rate has been discussed.}. The isotopes are assumed to be accelerated
in a decay ring, where they coincide with electrons accelerated at the same $\gamma$ 
factor and a boosted neutrino beam is produced towards the detector. At the distance 
of the baseline L the neutrinos hit the detector at a radial distance R from the 
beam axis and their energy in the laboratory frame (rest frame of the detector) can be 
expressed as
\begin{eqnarray}
E_{\nu}(R)=\frac{Q}{\gamma}\left[1-\frac{\beta}{\sqrt{1+(R/L)^2}}
\right]^{-1}\approx \, \frac{2\gamma Q}{1+(\gamma R/L)^2}.
\label{equ:Energy}
\end{eqnarray}
The derivation of this formula can be found in the Appendix, whereas the
approximation is taken from \cite{Sato:2005ma} and can be obtained in
the limit of large $\gamma$ with $\beta\approx 1-\frac{1}{2\gamma^2}$ and
$(R/L)\ll 1$. At the beam center the neutrino energy is maximal at a
value of $E_\nu=2\gamma Q$ and decreases for larger distances from the
beam center. Since the neutrino energy is a function of the distance from
the beam center, a position measurement within the detector allows a
precise reconstruction of the neutrino energy. We assume a Water
Cherenkov detector with a fiducial mass of 500~kt. The large detector mass
allows to collect enough statistics that is needed to gain from the superb
energy resolution and can have large geometrical size in order to have a
enough broad energy window, since the minimal measurable energy depends
on the maximal distance from the beam center. We assume the geometry of
the detector to be as shown in \figu{detector}. The radius of the
detector is set to $R_{\mathrm{max}}=100\, \mathrm{m}$ so that the depth is still 
approximately 64~m and a reconstruction of the Cherenkov rings and
electron/muon identification remains possible. The position measurement
should be optimized for this kind of experimental setup and reach at least a 
resolution of $\Delta R=$30~cm, which has been the estimated vertex resolution 
at \SK\ for fully-contained single ring events\cite{Ashie:2005ik}. Furthermore, 
the vertex resolution for muon events,~\ie the monobeam signal events, is 
slightly better than for electron events and can even reach a resolution 
of 25~cm in the energy window of interest. It should be mentioned that the very good position measurement
resolution can only be transfered into an excellent energy resolution if the systematical uncertainty in the beam spread can be reasonably
controlled. This means that the divergence of the stored isotopes perpendicular to the beam line must satisfy the condition
$p_x/p_z\lesssim \Delta R/L$ before the decay. Otherwise, the superb energy resolution that is assumed in this work could not be achieved
although the position measurement is accurate. This translates for baselines that are discussed in the following into the
requirement of a beam divergence $p_x/p_z\lesssim 1\ \mu \mathrm{rad}$ for the mother nuclei in the storage ring and seems hardly
feasible. However, it should be noted that beam divergences of $\mathcal{O}(1\ \mu \mathrm{rad})$ are already discussed, for 
instance for the proton beam of the LHC for the operation of the TOTEM experiment \cite{Deile:2004gt}.

For having neutrino energies beyond the Cherenkov threshold and allow for
electron/muon discrimination, we only discuss monobeam setups with
neutrino energies above 400 MeV. The signal efficiencies and background
rejection factors follow the analysis from \cite{Burguet-Castell:2003vv}
(mainly the low gamma beta beam therein). Above 400 MeV up to 700 MeV
there was found a signal efficiency of approximately 0.55 for the appearance
measurement of muon neutrinos, which we take to be the signal efficiency 
of the discussed monobeam scenarios. Although the signal efficiency in
\cite{Burguet-Castell:2003vv} decreases slightly for higher neutrino
energies, we assume the signal efficiency to stay stable up to the
highest energies discussed in this work ($\mathrm{E_\nu\lesssim 1.4 GeV}$), since
the monobeam setup does not rely on the quasi-elastic events only, because
the energy reconstruction can be performed by the position measurement 
within the detector. The rejection factors for background coming from 
neutral current events with single pion production at energies above 400 MeV 
are found to be below $10^{-3}$ in \cite{Burguet-Castell:2003vv}, whereas we 
assume this background rejection to be at a level of $10^{-4}$. This is a 
quite optimistic assumption, and it is not clear, that this could be achieved. 
However, note that the mismatch of ordinary energy reconstruction and energy 
reconstruction by position measurement due to the carried away missing energy 
of the neutrino in neutral current reactions could give a further rejection ability of such 
background events. We assume a systematical uncertainty of 2.5\% for the 
signal events and 5\% for the background events, as also assumed for the 
typical beta beam scenarios,~\ie in \cite{Huber:2005jk}.  The uncertainty of the 
signal events has probably to be called optimistic, but since we will find 
that the main effect will come from correlations and degeneracies \cite{Koike:2000jf,Minakata:2001qm,Barger:2001yr}, 
a value of 5\% would not have much impact to the results of this work. 

As can be understood from \equ{Energy}, the energy window of the analysis is, due to the
technique of energy reconstruction, limited by the size of the detector to the interval

\begin{eqnarray}
\frac{2\gamma Q}{1+(\gamma R_{\mathrm{max}}/L)^2} \, \le \, E_{\nu} \, \le \, 2\gamma Q, 
\label{equ:window}
\end{eqnarray}

so that the energy window is completely fixed after the baseline L and the acceleration 
factor of the ions $\gamma$ is chosen. So finding an optimal Setup is more complicated 
as it is for example in the case of beta beams, since choosing a perfect pair of L 
and $\gamma$ to exactly measure at the first oscillation maximum can suffer from 
an energy window that is to small to allow resolving correlations and degeneracies.
However, adjusting the baseline to smaller baselines in order to have a lower minimal 
energy also shifts the oscillation maximum to lower energies, while going to higher 
values of $\gamma$ not only shifts the maximal energy but also the minimal energy to 
higher values. So, the whole energy window moves away from the oscillation maximum 
although it is broadened. Therefore, in the next sections we discuss the potential 
and performance of the following different reference scenarios of monobeam setups: 

\begin{itemize}
\item {\bf Setup~I}: The Water Cherenkov detector with a fiducial mass of 500~kt 
is located at a baseline of L=600km, the mother nuclei $^{110}_{50}\mathrm{Sn}$ 
are accelerated with $\gamma=2500$ and 10 years of data taking are assumed at
the number of $10^{18}$ electron capture decays per year.

\item {\bf Setup~II}: The Water Cherenkov detector with a fiducial mass of 500~kt 
is located at a baseline of L=250km, the mother nuclei $^{110}_{50}\mathrm{Sn}$ 
are accelerated with $\gamma=2000$ and 10 years of data taking are assumed at
the number of $10^{18}$ electron capture decays per year

\item {\bf Setup~III}: The Water Cherenkov detector with a fiducial mass of 500~kt 
is located at a baseline of L=600km, the mother nuclei $^{110}_{50}\mathrm{Sn}$ 
are accelerated with $\gamma=900$ and $\gamma=2500$ consecutively, and 5 years 
of data taking are assumed in each of the two phases so that as for Setup~I and 
II the total running time is 10 years. The number of $10^{18}$ electron capture 
decays per year is assumed for both phases.
\end{itemize}
 
\begin{figure}[t]
\begin{center}
\includegraphics[width=0.45\textwidth]{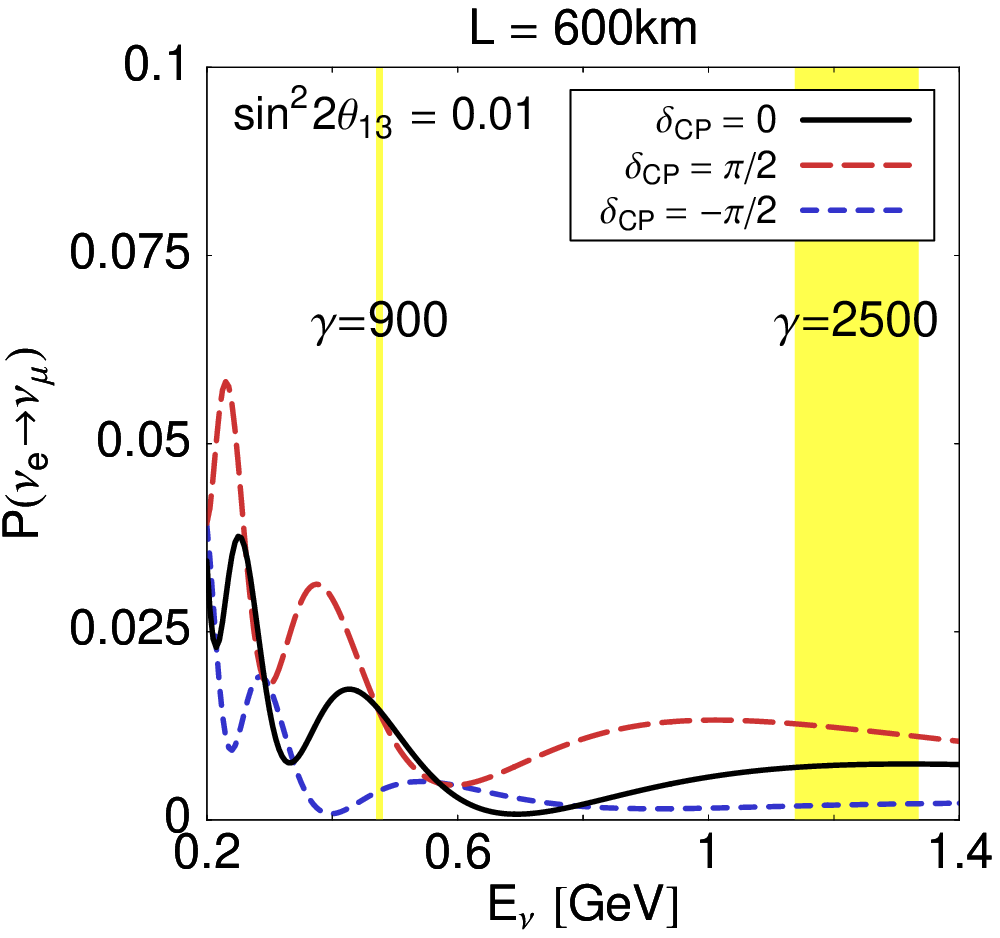} \hspace{0.3cm}
\includegraphics[width=0.45\textwidth]{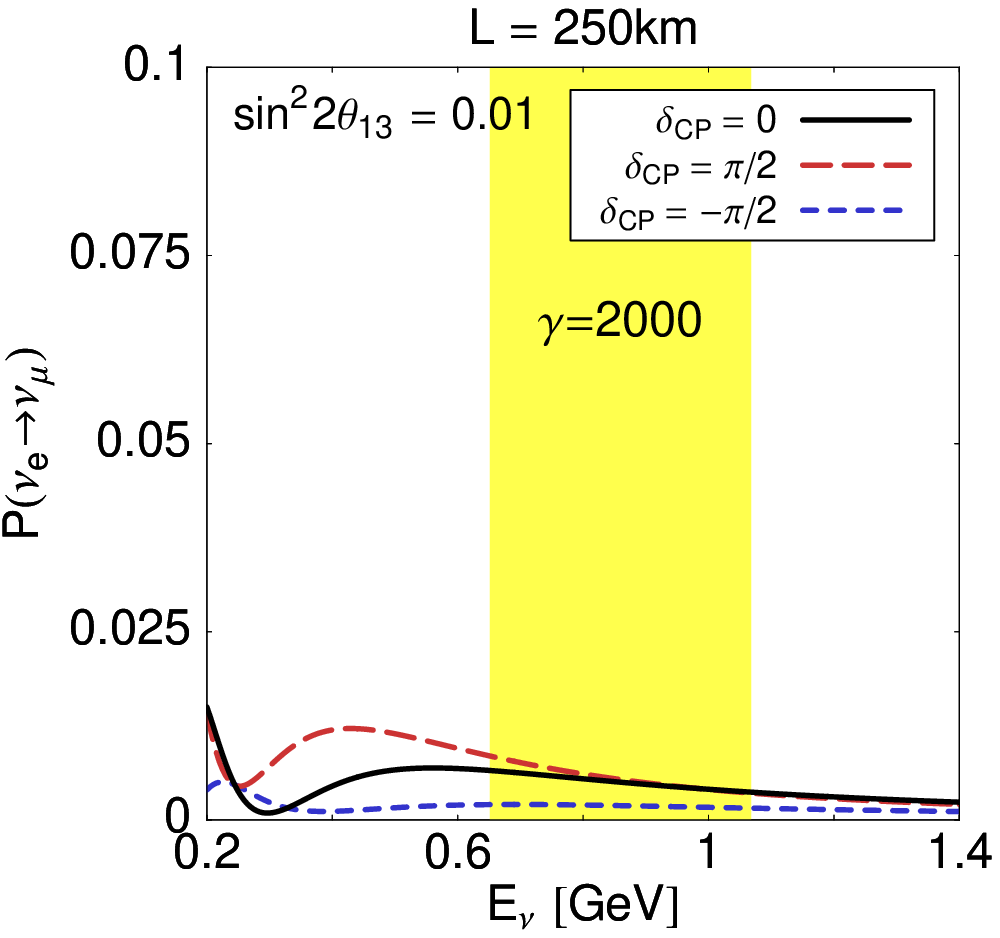} \hfill
\end{center}
\mycaption{\label{fig:probability} The appearance probability P($\nu_e\rightarrow\nu_\mu$) 
as a function of the neutrino energy $\mathrm{E_\nu}$ at a baseline of L=250km (left-hand
side) and L=600km (right-hand side). The oscillation parameter values are the ones 
from \equ{params}, $\sin^22\theta=0.01$ and the values for the phase $\deltacp$ are chosen 
as labeled in the plot legend. The vertical yellow (grey) bands indicate the energy window 
of the analysis that is given for a detector radius of $R_{\mathrm{max}}=100\mathrm{m}$, the 
respective baseline L and the chosen $\gamma$ factors,~\ie $\gamma=900$/$\gamma=2500$ 
for the scenarios at L=600km (Setup~I and Setup~III) and $\gamma=2000$ for the scenario at 
L=250km (Setup~II).}
\end{figure}

Setup~I is located at the first oscillation maximum, but the energy window is not very 
broad compared to the width oft the oscillation maximum peak, therefore we also discuss the 
second scenario, Setup~II, with a broader energy window which on the other hand is 
located slightly off the first oscillation maximum at higher neutrino energies due to 
the smaller baseline. Then again, because of the smaller baseline higher event rates can be
obtained at Setup~II. With Setup~III we discuss the potential in resolving the correlations 
and degeneracies with a monobeam experiment by a combination of data from the first 
oscillation maximum and also the second oscillation maximum. This combination should be a
powerful tool to resolve the degeneracies and the importance of the second oscillation maximum
has been discussed in~\cite{Arafune:1997hd}. Since the first oscillation maximum phase
at Setup~III is comparable to Setup~I, the gain from the additional 
measurement at the second oscillation maximum can directly be read off the comparison 
of Setup~I and Setup~III. The exact width of the corresponding energy windows of the 
setups and their location respectively to the oscillation maxima are shown in \figu{probability}. 
Note, that Setup~III makes use of the combination of different $\gamma$ which was also the strategy for the
``high $Q$-low $\gamma$'' electron capture beam experiment scenarios as discussed in \cite{Bernabeu:2005jh,Lindroos:2005jz,Bernabeu:2005ks,
Bernabeu:2005kq,Bernabeu:2005zs}. However, for these scenarios this strategy was required to obtain spectral information at the first
oscillation maximum, while Setup~III provides spectral information at the higher $\gamma=2500$ and information from the second
oscillation maximum is included with the second arrangement of $\gamma=900$. This can be seen in \figu{probability}.  

There, the appearance probability $\mathrm{P(\nu_e\rightarrow\nu_\mu)}$ is plotted for $\stheta=0.01$ 
and three choices of $\deltacp$ (the other oscillation parameters are chosen as in 
\equ{params}). The yellow (grey) bands indicate the energy window of the analysis for 
Setup~I and III in the left-hand side and Setup~II in the right-hand side. It can be 
seen that the energy window for the choice of L=600km and $\gamma=900$ is essentially 
only a very narrow band while for the higher values of $\gamma$ indeed a broader energy 
window can be covered over the whole radius of the detector. However, the energy window 
of Setup~I is too narrow to cover the first oscillation maximum for the different choices 
of $\deltacp$. For $\deltacp=0$ the peak of the first oscillation maximum lies inside the 
energy window of the analysis but for the maximally CP violating values for $\deltacp$ the 
peak moves outside the energy window. The energy window of Setup~II lies above the first 
oscillation maximum independent of $\deltacp$ but we will show in the next sections that 
Setup~I will suffer more from correlations and degeneracies than Setup~II since the latter 
benefits from a higher event rate due to the smaller baseline and the larger energy window, 
where the superb energy resolution can evolve. 

Note, that the number of electron capture decays per year taken for the reference 
scenarios is of the order of the ``high Q'' electron capture scenarios discussed 
in~\cite{Bernabeu:2005jh, Lindroos:2005jz,Bernabeu:2005ks,Bernabeu:2005kq, 
Bernabeu:2005zs} and also the order of beta decays per year discussed for the beta beam 
scenarios~\cite{Zucchelli:2002sa,Mezzetto:2003ub,Bouchez:2003fy,Mezzetto:2004gs,
Burguet-Castell:2003vv,Burguet-Castell:2005pa,Terranova:2004hu,Donini:2004hu,
Donini:2004iv,Donini:2006dx,Donini:2006tt,Huber:2005jk,Campagne:2006yx}. However, 
because of the long lifetime of the $^{110}_{50}\mathrm{Sn}$ electron capture of 
4.11~h in the rest frame the feasibility to achieve a number of $10^{18}$ per year 
seems out of range if it cannot be enhanced. This enhancement of the electron 
capture processes could be achieved by a laser irradiation as discussed 
in~\cite{Ikeda:2005yh,Yoshimura:2005zq,Nomura:2006ik}. Furthermore, as is also the 
case for high gamma beta beams~\cite{Burguet-Castell:2003vv,Huber:2005jk} the high 
gamma values require a very large accelerator complex of the size of the \mbox{\sf LHC}. 

For reasons of comparison and to put the performance of 
the monobeam setups into perspective we will compare the results to a standard neutrino 
factory setup with a 50~kt MID detector at a baseline of L=3000km and a parent energy of 
the stored muons $E_\mu=50\mathrm{GeV}$. This neutrino factory setup is similar to the standard scenario
for a neutrino experiment \cite{Cervera:2000kp}, commonly known as NuFact-II, with $1.06\cdot10^{21}$ useful muon decays per 
year (corresponding to $5.3\cdot10^{20}$ useful muon decays 
per year and polarity for a simultaneous operation with both polarities).
The details of the neutrino factory experiment description follow the description of the NuFact-II scenario in \cite{Huber:2002mx}. 
We assume a runtime of five years in each polarity, so that the 
total running time is 10 years as for the discussed monobeam setups.
Furthermore we will also consider an optimized neutrino factory scenario at the end of Section 4, where compared to the
standard neutrino factory scenario a second detector similar to the standard detector at L=3000km is installed at the magic baseline
L=7500km.\footnote{The optimized scenario furthermore uses an optimized disappearance channel with the MINOS energy threshold
\cite{Ables:1995wq} while the muon
CID with the implied CID cut threshold is only used for the golden appearance channel. See \cite{Huber:2006wb} for details.} 
   
The analysis throughout this work is performed with the \globes software \cite{Huber:2004ka,
Rolinec:2006kp} and the incorporated Poisson $\chi^2$-analysis. Details can be found in 
the \globes manual~\cite{GLOBES}. Since the monobeam only measures $\nu_\mu$-appearance and 
could additionally only observe $\nu_e$-disappearance, the leading atmospheric parameters
$\sin^22\theta_{23}$ and $|\ldm|$ cannot be determined as would be the case at a neutrino 
factory with a measurement in the $\nu_\mu$-disappearance channel. Thus, correlations with the
leading atmospheric parameters would spoil the potential of the monobeam experiment alone, 
as also would be the case for a beta beam for the same reasons. Therfore, we adopt the same 
technique as in \cite{Huber:2005jk} and add the $\nu_\mu$-disappearance information from a 
simulation of the superbeam experiment \JHFSK. The corresponding appearance information 
is excluded, so that information on $\stheta$ and $\deltacp$ is solely collected by
the monobeam experiment (see~\cite{Huber:2005jk} for details). The errors on the solar parameters
are taken to be 5\% on each, $\Delta m^2_{21}$ and $\theta_{12}$.

As input or so-called true values within the simulations, we use, unless stated otherwise 
the following parameter values, close to the current best fit values 
(see~\Refs~\cite{Fogli:2003th,Maltoni:2004ei,Bahcall:2004ut,Bandyopadhyay:2004da}):
\begin{eqnarray}
\ldm=2.5\cdot10^{-3}\,\mathrm{eV}^2\quad\sin^22\theta_{23}=1.0 \, , \nonumber \\
\Delta m^2_{21}=8.2\cdot10^{-5}\,\mathrm{eV}^2\quad\sin^22\theta_{12}=0.83. 
\label{equ:params}
\end{eqnarray}
Note, that the octant-degeneracy \cite{Fogli:1996pv} does not affect our results, 
since we choose $\sin^22\theta_{23}$ to be maximal and thus the octant-degenerate 
solution appears at the same point in the parameter space as the original solution 
in the parameter space. So, if it is stated that effects of degeneracies are taken 
into account in the analysis, only the intrinsic 
$\stheta$-$\deltacp$-degeneracy~\cite{Burguet-Castell:2001ez} and the 
sign($\ldm$)-degeneracy~\cite{Minakata:2001qm} are regarded out of the complete set 
of the so-called eight-fold degeneracy~\cite{Barger:2001yr}.

\section{Sensitivity to $\boldsymbol{\stheta}$}

\begin{figure}[t!]
\begin{center}
\includegraphics[width=0.4\textwidth]{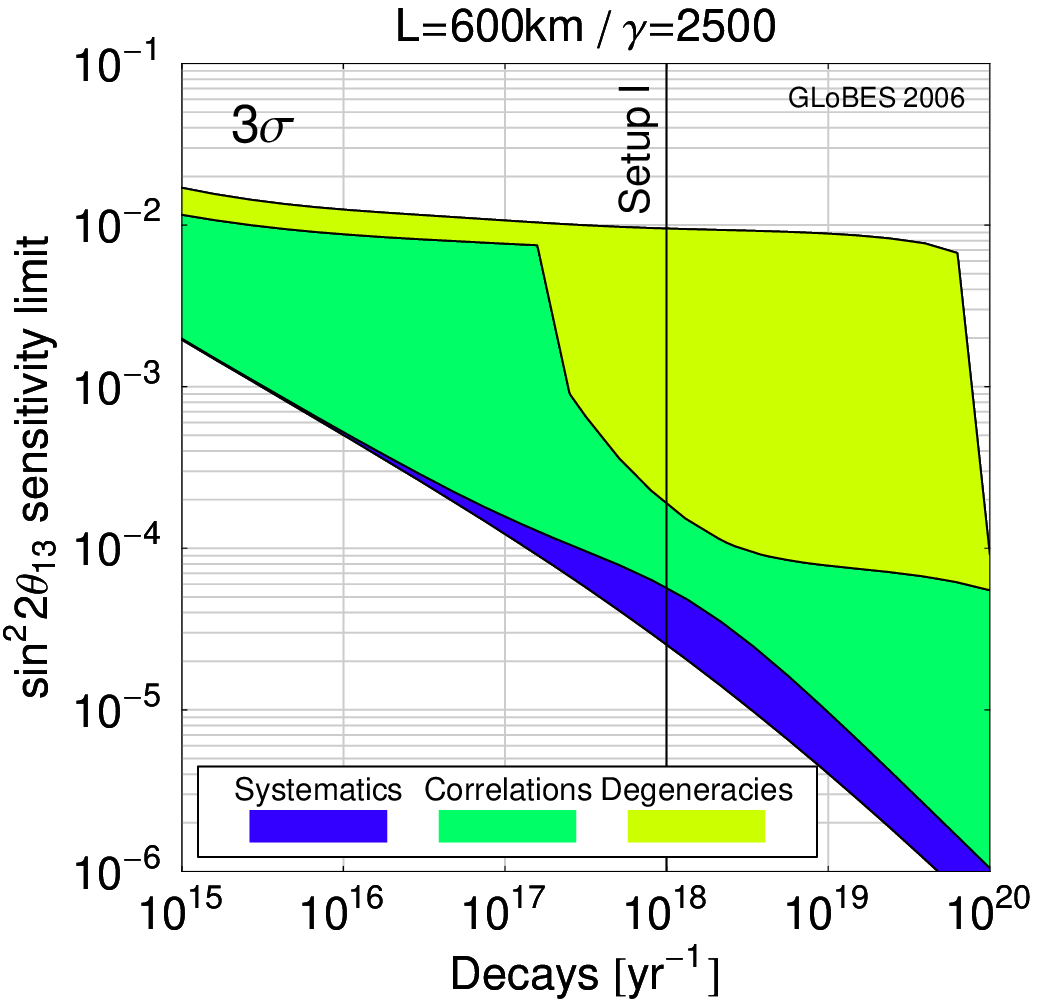} \hspace{0.6cm}
\includegraphics[width=0.4\textwidth]{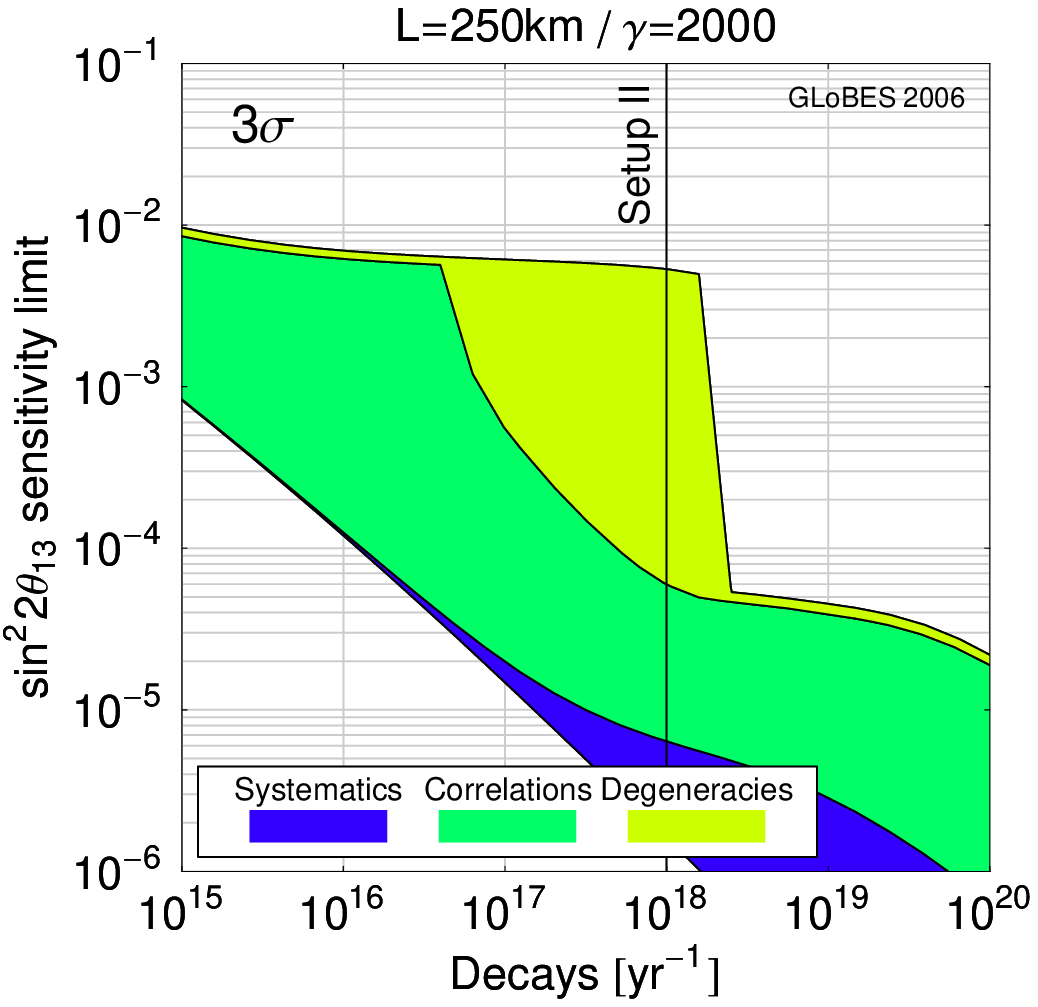} \\
\includegraphics[width=0.4\textwidth]{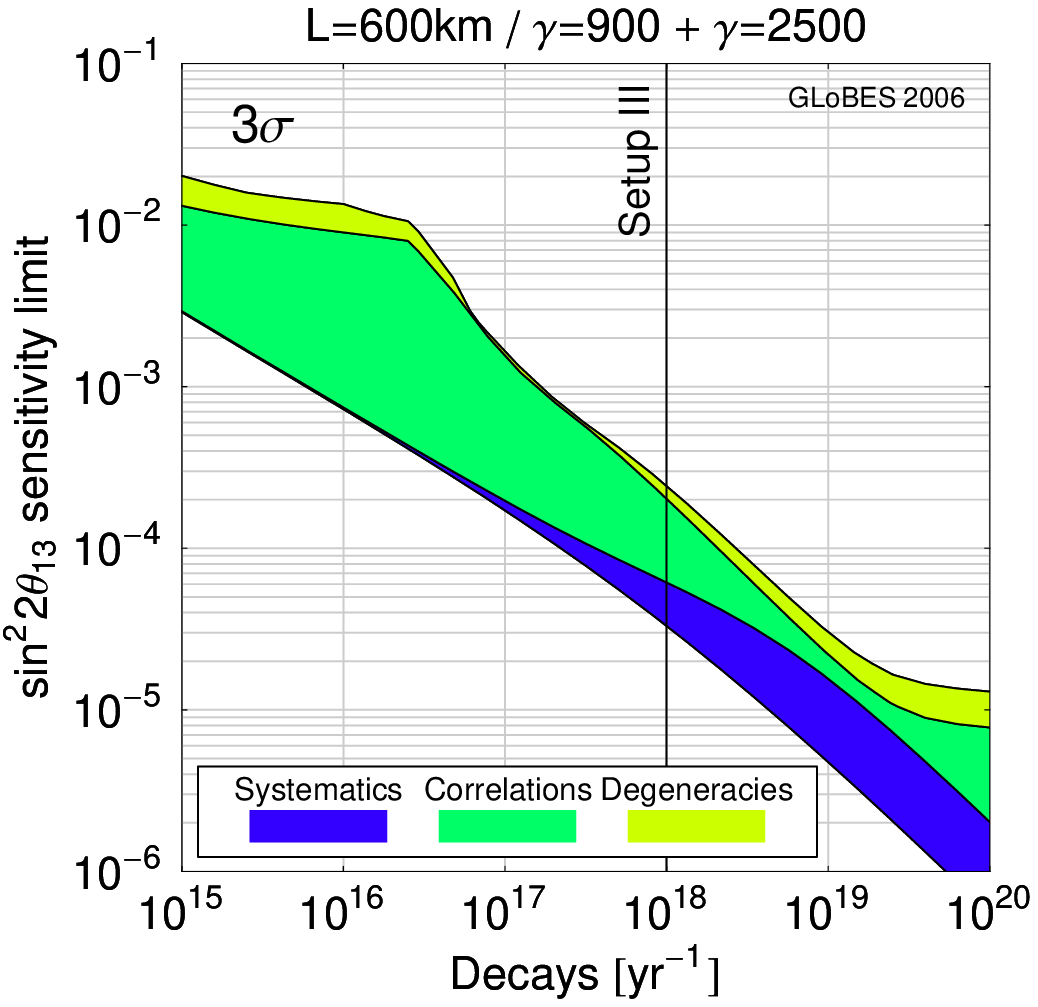} 
\end{center}
\mycaption{\label{fig:decays} The sensitivity to $\stheta$ at the 3$\sigma$ 
confidence level for the monobeam scenarios L=600km/$\gamma=2500$, L=250km/$\gamma=2000$, 
and L=600km/$\gamma=900+\gamma=2500$ as a function of the number of decaying ions per 
year including statistics, systematics, correlations, and degeneracies. The lowest 
curve represents the pure statistical sensitivity limit to $\stheta$ and the colored 
bands indicate the effect of switching on systematics (blue/dark grey), correlations 
(green/middle grey), and degneracies (yellow/bright grey) so that the final 
sensitivity limit is given by the upper curve.}
\end{figure}

The sensitivity to $\stheta$ is calculated under the hypothesis of true $\stheta=0$. 
The sensitivity limit at a certain confidence level is then the maximal fit value of 
$\stheta$ that still fits the simulated data at the chosen confidence level,~\ie it 
would be the lower bound to $\stheta$ that the experiment could achieve in case of 
vanishing true $\stheta$. It is well known, that the main problem is to resolve the 
correlations with the other oscillation parameters and the so-called eight-fold
degeneracy. In \figu{decays} the sensitivity to $\stheta$ is shown at the $3\sigma$ 
confidence level as a function of the number of decays per year for the monobeam 
scenarios at L=600km/$\gamma=2500$, L=250km/$\gamma=2000$, and L=600km/$\gamma=900+\gamma=2500$. 
The vertical lines indicate the reference setups at a number of $10^{18}$ ion decays 
per year. 
In each plot the lowest curve represents the pure statistical limit to $\theta$ 
and the colored bands show how the sensitivity degrades if also systematics (blue/dark 
grey band), correlations (green/middle grey band), and degeneracies (yellow/bright grey 
band) are taken into account. The final achievable sensitivity limit to $\stheta$ is given 
by the upper curve. Obviously the statistical and systematical sensitivity limit to 
$\stheta$ at all three scenarios in \figu{decays} can reach to very small values of 
$\stheta$ due to the very large statistics in the Water Cherenkov detector. However, the 
monobeam scenario at L=600km/$\gamma=2500$ can resolve the correlations not until an 
exposure of $10^{17}$ decays per year. The point where the degeneracies can be resolved 
is reached not until approximately $10^{20}$ decays per year, which of course is beyond 
any feasibility. So despite the improvement of the statistical limit with higher exposures 
the final sensitivity limit to $\stheta$ stays relatively stable a approximately 
$\stheta\approx10^{-2}$ independent of the number of decays per year. The monobeam scenario 
at a baseline of L=250km and $\gamma=2000$ suffers from the same problem. First, the 
sensitivity limit does only slightly improve and almost stays stable. Beyond exposures 
of $10^{18}$ decays per year this scenario starts to resolve the degeneracies and the 
sensitivity limit to $\stheta$ improves significantly. From \figu{decays} it becomes 
clear, that the technique of a high gamma monobeam with its superb energy resolution 
in a narrow energy window is not able to resolve the correlations and degeneracies in 
a measurement at just one $\gamma$. The scenario at a baseline of L=600km allows to 
measure in the second oscillation maximum since for L=600km this maximum is located 
above the Cherenkov threshold and events can be collected. The lower plot of \figu{decays} 
shows the sensitivity limit to $\stheta$ for such a scenario, where 5 years data taking
at $\gamma=900$ and 5 years data taking at $\gamma=2500$ is combined. Now, the correlations 
and degeneracies can be already resolved for lower exposures. We checked that it is not 
necessary to split up the two data taking phases into an equal period of five years 
each. The ability to resolve the correlations and degeneracies still remains if only 
2 years data taking at $\gamma=900$ are combined with 8 years at $\gamma=2500$
and the final sensitivity would be even slightly better since then more statistics could 
be collected at the first oscillation maximum. 

\begin{figure}[t]
\begin{center}
\includegraphics[height=0.4\textwidth]{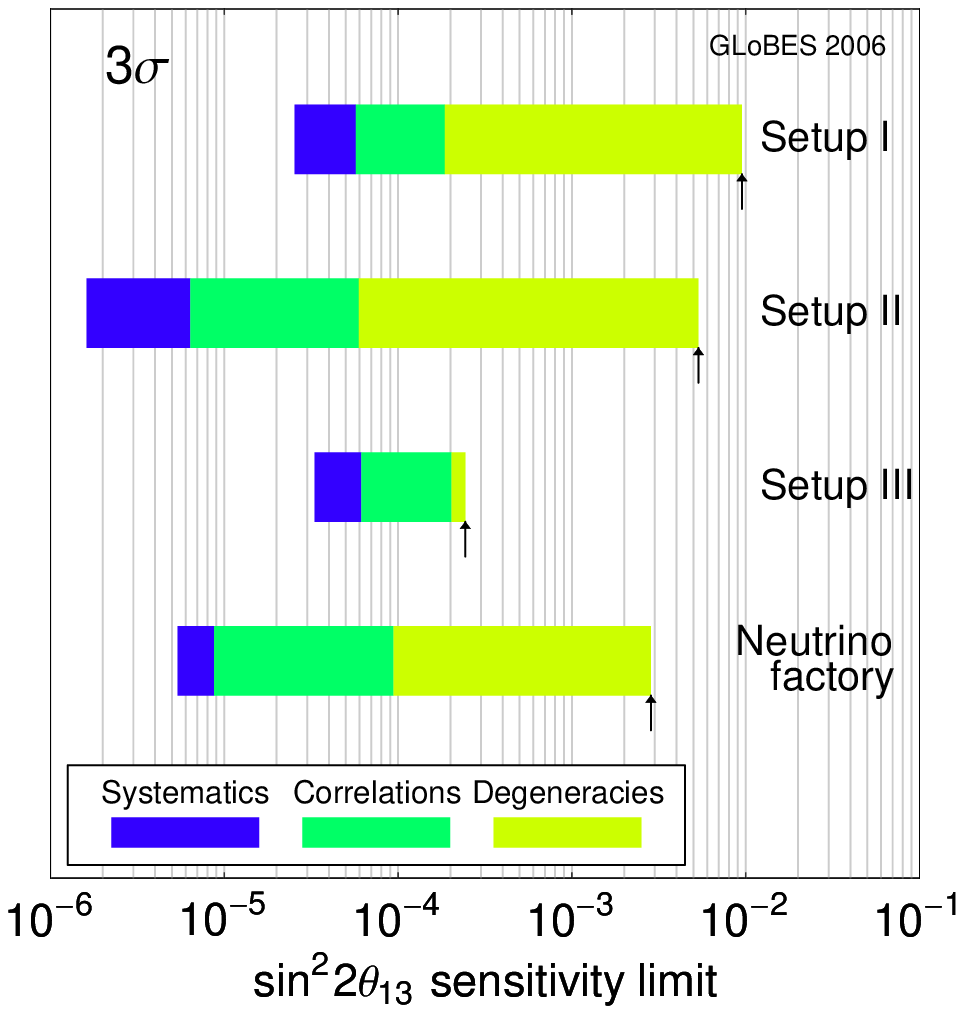} \hspace{0.6cm}
\includegraphics[height=0.4\textwidth]{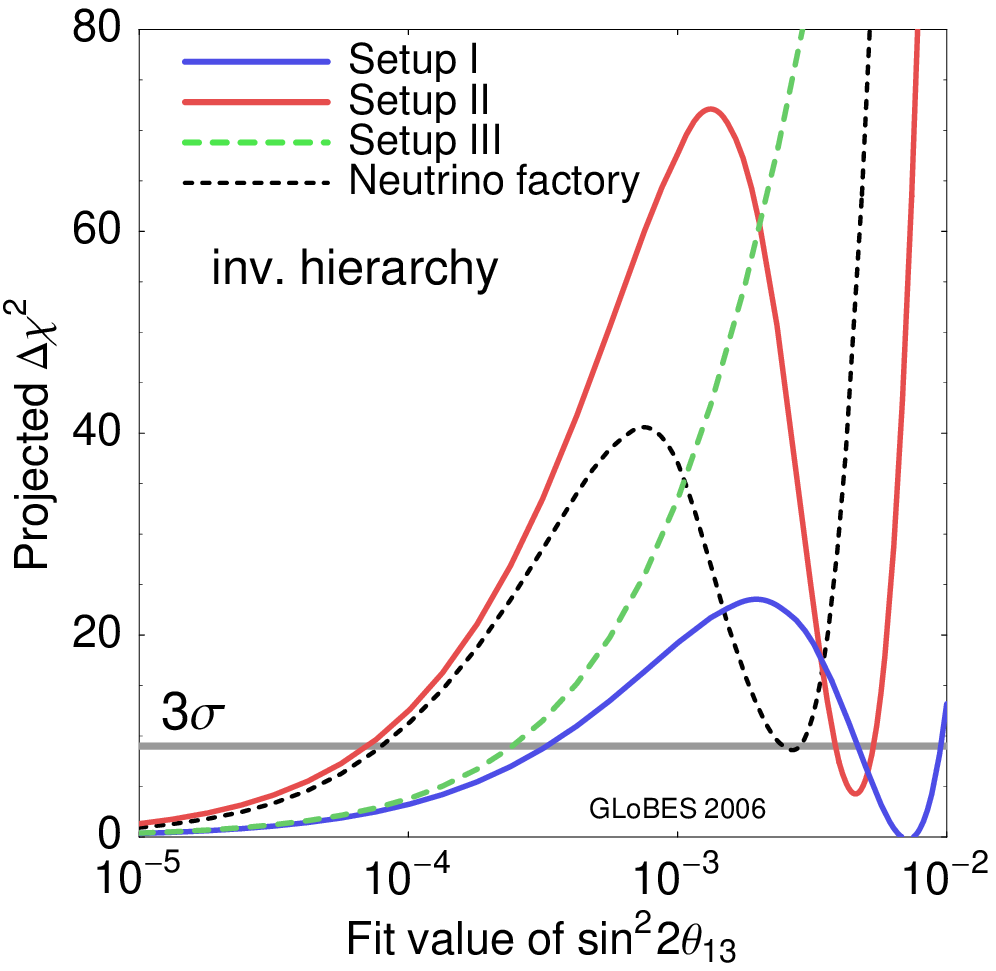}
\end{center}
\mycaption{\label{fig:SensComp} Left-hand side: Comparison of the sensitivity 
to $\stheta$ at the 3$\sigma$ confidence level at the monobeam scenarios 
Setup~I, Setup~II, Setup~II, and a neutrino factory at a baseline of L=3000km 
including systematics, correlations and degeneracies. The left edge of the bars 
indicates a pure statistical sensitivity limit. The right edges of the bars 
indicate the sensitivity limit after switching on systematics (blue/dark grey), 
correlations (green/middle grey), and correlations (yellow/bright grey), so 
that the rightmost edge gives the final 
sensitivity limit to $\stheta$. Right-hand side: The projected $\Delta \chi^2$ 
as a function of the fit value of $\stheta$ fitted under the assumption of 
inverted hierarchy while the true values are given with $\stheta$ and normal 
hierarchy. The rightmost intersections of the curves with the grey horizontal 
$3\sigma$ line give the right edges of the bars in the plot on the left-hand side.}
\end{figure}

For reasons of comparison, the sensitivity 
to $\stheta$ at Setup~I, Setup~II, and Setup~III are again shown in the left-hand side 
of \figu{SensComp} and confronted with the sensitivity limit obtainable at the standard 
neutrino factory scenario. The neutrino factory also suffers from the correlations and 
degeneracies. But as can be seen in the right-hand side of \figu{SensComp}
the difference is that the neutrino factory can almost resolve the degenerate solution. 
There, the projected $\Delta \chi^2$ is shown as a function of the fit value of 
$\stheta$ for the degenerate solution with the wrong sign,~\ie inverted hierarchy 
while the positive $\ldm$ was taken as input true value. The degenerate solution 
appears for the neutrino factory scenario at a $\Delta \chi^2$ only slightly below the
3$\sigma$, while the degenerate solution for Setup~I appears at $\Delta \chi^2=0$ 
and thus fits as good as $\stheta=0$. On the other hand, with Setup~III there does 
not appear a second local minimum in the projected $\Delta \chi^2$ so the combination 
of first and second oscillation maximum data gives a strong tool to resolve the 
degeneracy. However, resolving the degeneracies remains the main problem if one want 
to reach to very small values of $\stheta$ and one could also think of a combination 
of a monobeam setups with the anti-neutrino running of a standard beta beam scenario. 
It should be noted that the performance of a neutrino factory could be improved by 
additional data from the silver channel $\nu_\mu\rightarrow\nu_\tau$ \cite{Donini:2002rm,Autiero:2003fu}, a second 
detector at the magic baseline \cite{Barger:2001yr,Lipari:1999wy,Huber:2003ak} or a lower threshold (see~\cite{Huber:2006wb}).

\section{Sensitivity to CP violation}

\begin{figure}[t!]
\begin{center}
\includegraphics[width=0.85\textwidth]{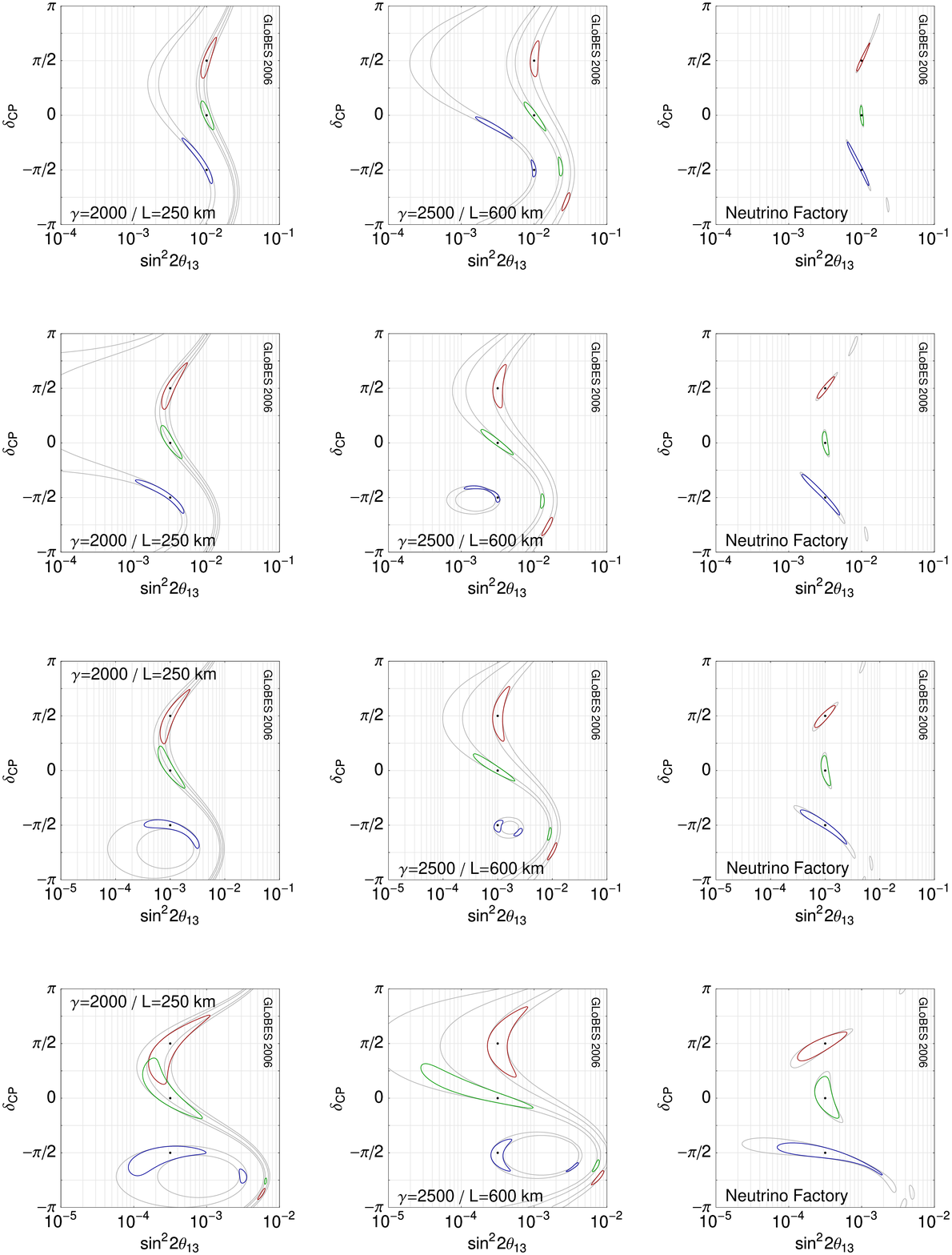}
\end{center}
\mycaption{\label{fig:regions} The allowed regions in the $\stheta$-$\deltacp$-plane 
for the true values indicated by the black dots at 3$\sigma$ for Setup~I (left
column), Setup~II (middle column), and a standard neutrino factory (right column) 
for reasons of comparison. Only the correlation between $\stheta$ and $\deltacp$ 
are taken into account and all other parameters are fixed to values of \equ{params}.
The plots also contain the allowed regions at 3$\sigma$ for total rates only 
(grey solid lines).}
\end{figure}
Due to the continuous intrinsic $\stheta$-$\deltacp$-degeneracy a total
rates analysis of appearance data of neutrinos only would give continuous
bands as allowed regions in the $\stheta$-$\deltacp$ plane. If combined
with a second band from appearance data of anti-neutrinos only two
intersections, the true and the degenerate allowed region remain. Adding
the spectral information obtained with conventional energy resolution,
the degenerate solution can be resolved in most cases. This is the
planned procedure at superbeam experiments, neutrino factories as well
as beta beam experiments to resolve the $\stheta$-$\deltacp$-degeneracy. 
However, at a monobeam experiment only neutrino appearance is observable and
the question arises, if and under which circumstances the superb energy
resolution abilities of a monobeam could in principle compete in
resolving the $\stheta$-$\deltacp$-degeneracy. Since we found in the
last section that the ability in resolving the degeneracies does not
appear until a large number of decays per year, we will fix this value to
$10^{18}$ decays per year in all the following considerations and only 
discuss the fixed scenarios Setup~I, Setup~II, and Setup~II. In
\figu{regions} the allowed regions in the $\stheta$-$\deltacp$-plane at
the 3$\sigma$ confidence level are shown for different choices of input
true values. This figure is for illustrative purposes only and no
correlations with the other oscillation parameters is
considered,~\ie they are kept fixed to the values of
\equ{params}. The left column is for Setup~I (L=600km/$\gamma=2500$), 
the middle column is for the Setup~II (L=250km/$\gamma=2000$), 
and the right column shows the allowed regions obtained for
the standard neutrino factory setup for reasons of comparison. The
bands, indicated by the solid grey lines, represent the corresponding 
allowed regions at the 3$\sigma$ confidence level if only total rates are
considered. As expected, the total rates allowed regions for the
monobeam scenarios are bands that do not restrict $\deltacp$ at all
whereas for the neutrino factory already also the parameter space of
$\deltacp$ is restricted due to the information from neutrino and
anti-neutrino data. If spectral information is included to the analysis,
the neutrino factory allowed regions are not influenced significantly
and only the small degenerate solutions can be excluded, but for the
monobeam scenarios because of the superb energy resolution wide parts of
the bands can be excluded and only smaller allowed regions remain that
are comparable in size to the allowed regions from the neutrino factory
scenario. However, in some cases of choices of true values still
degenerate solutions remain. As mentioned before, we have ignored
correlations with the other oscillation parameters and also the
sign($\ldm$)-degeneracy here. In all of the further considerations, we
will focus on the sensitivity to CP violation if also these correlations
and all degeneracies are taken into account.

\begin{figure}[t!]
\begin{center}
\includegraphics[width=0.4\textwidth]{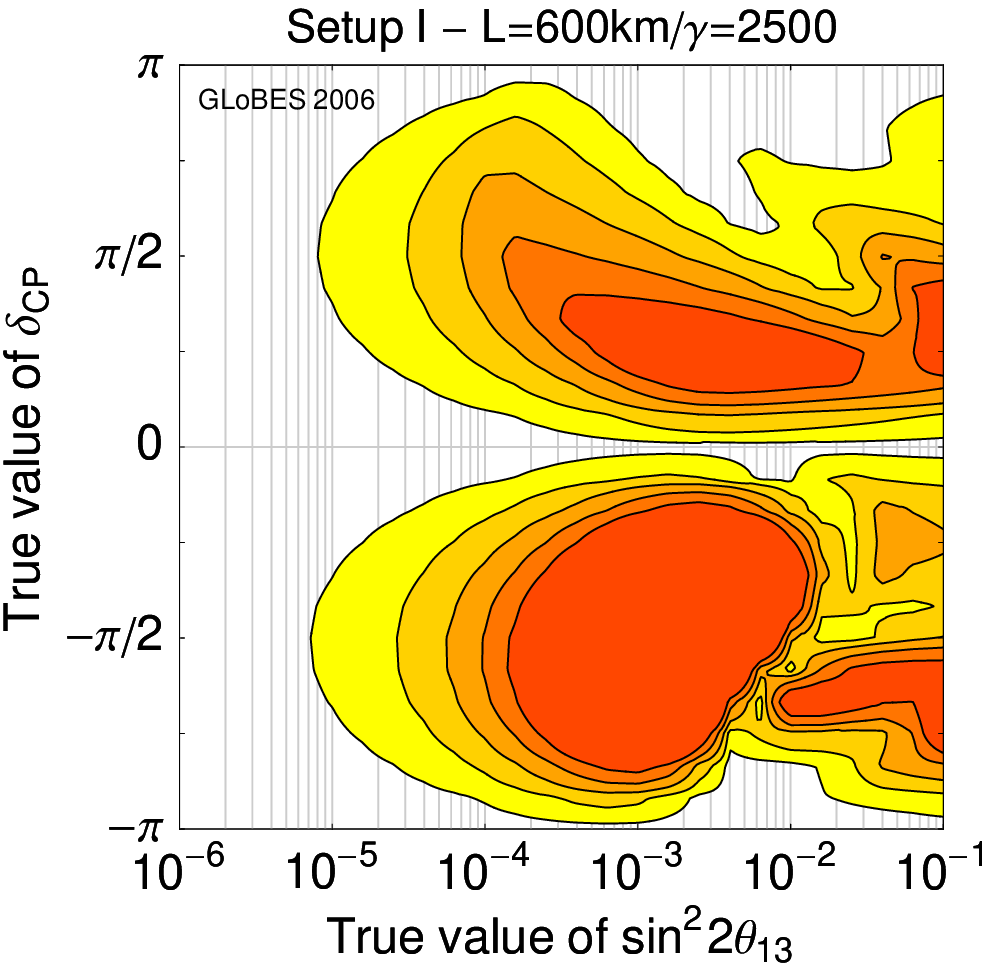} \hspace{0.6cm}
\includegraphics[width=0.4\textwidth]{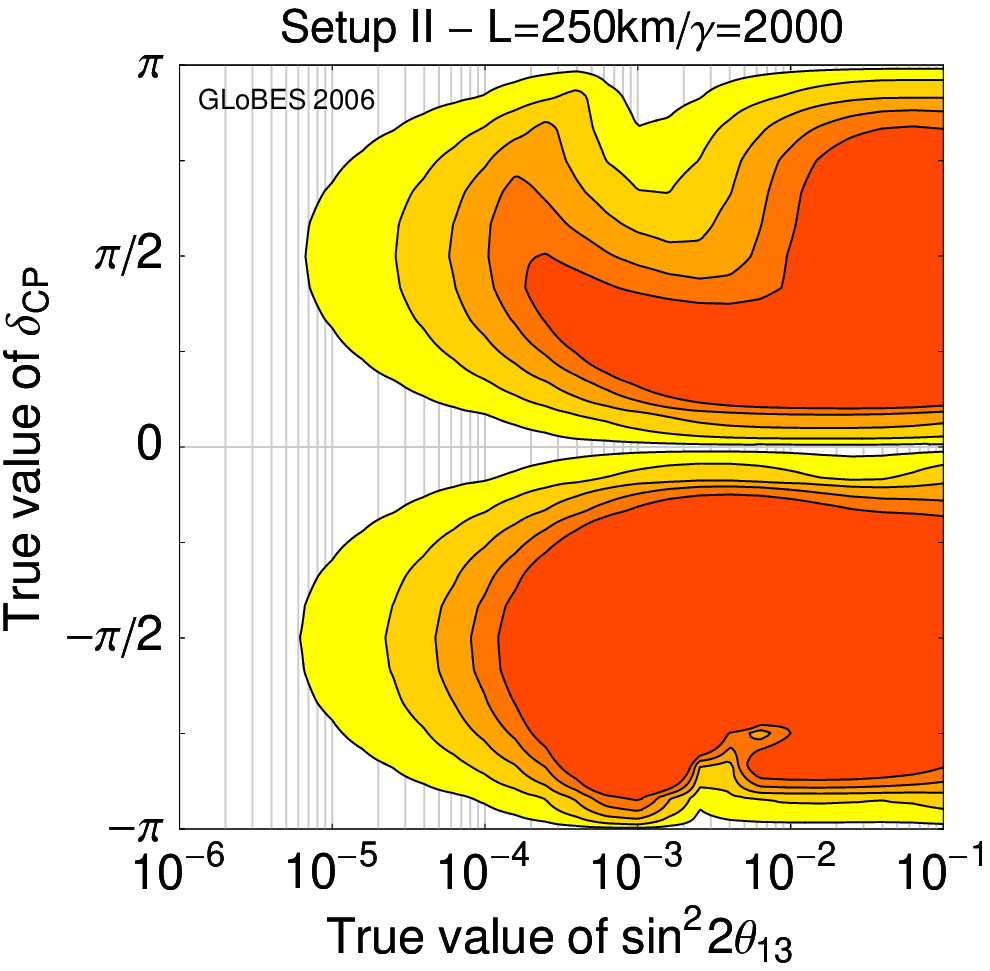} \\
\includegraphics[width=0.4\textwidth]{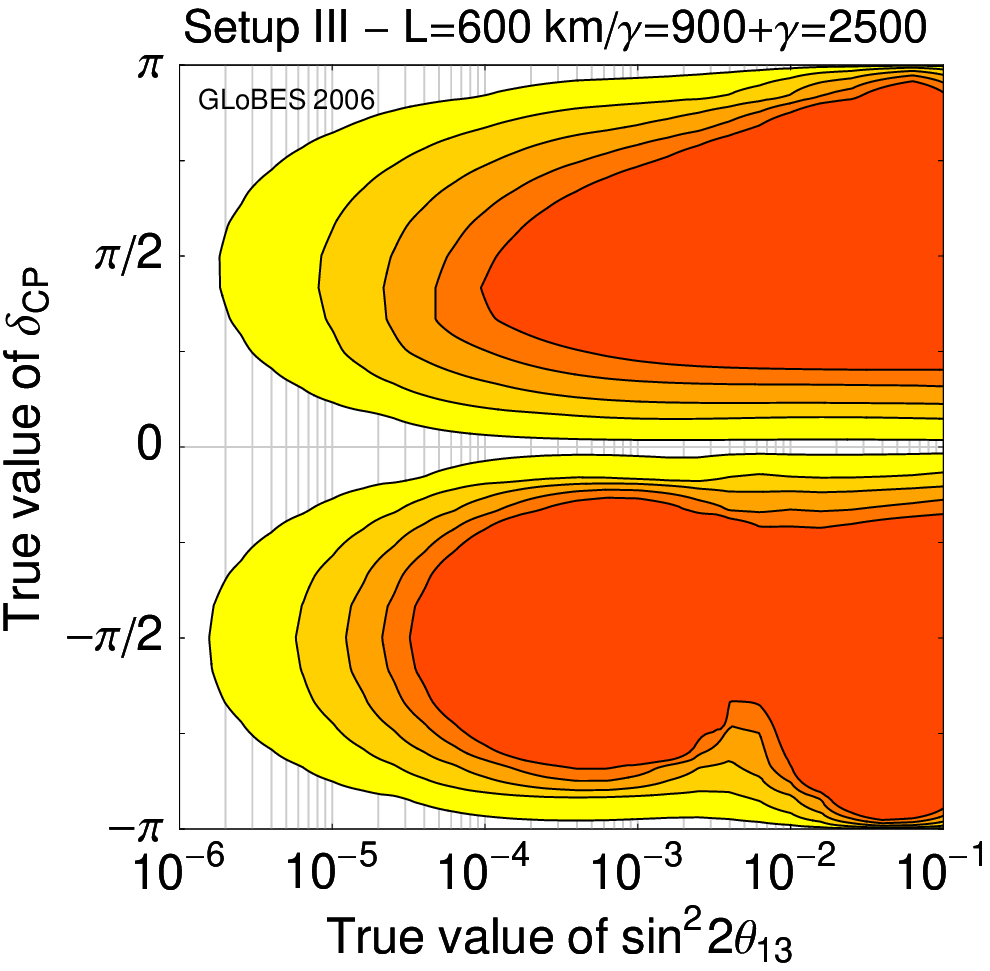} \hspace{0.6cm}
\includegraphics[width=0.4\textwidth]{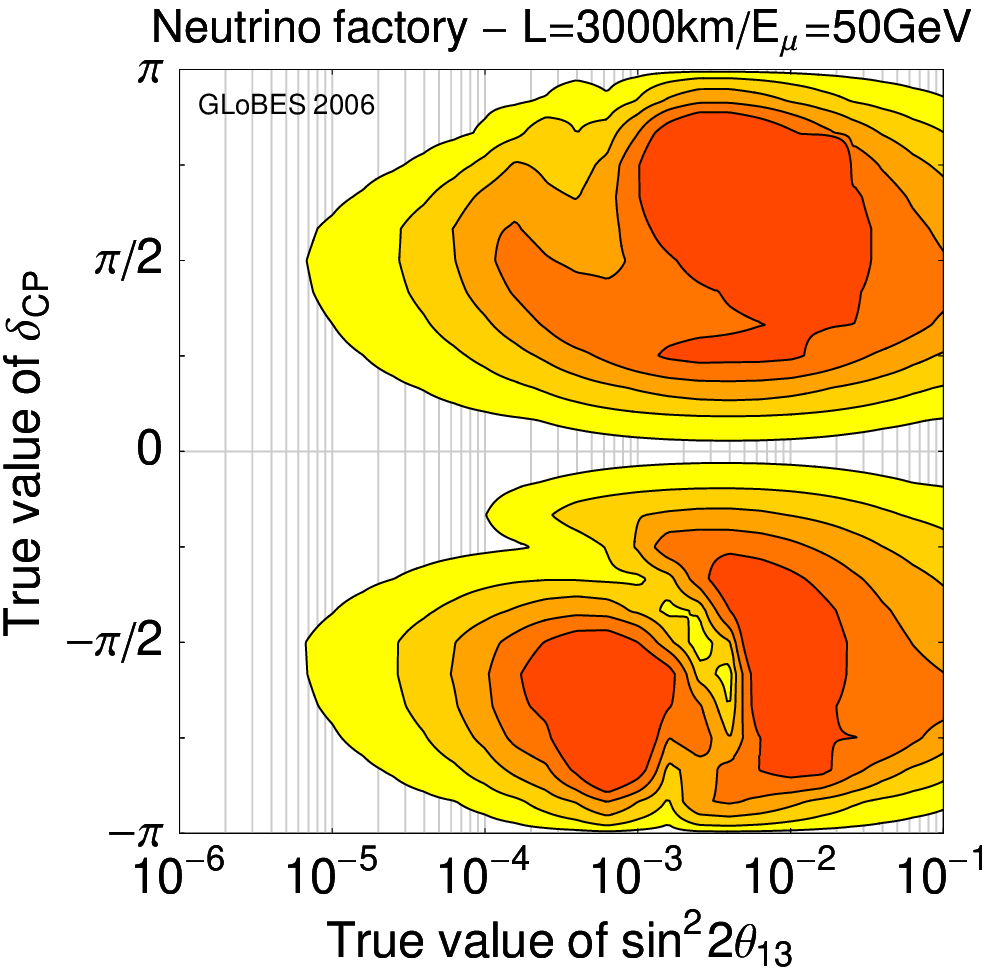}
\end{center}
\mycaption{\label{fig:anycp} Sensitivity to any CP violation at 1 (yellow/bright 
grey), 2, 3, 4, and 5$\sigma$ (red/dark grey) after 10 years of data taking as 
a function of the true values of $\stheta$ and $\deltacp$. The sensitivities are 
shown for the monobeam scenarios Setup~I (upper left-hand side plot), Setup~I 
(upper right-hand side plot), Setup~III (lower left-hand side plot) and a standard 
neutrino factory (lower right-hand side plot) for reasons of comparison. For a 
pair of true values within the shaded regions the CP conserving fit values 
$\deltacp=0$ and $\deltacp=\pi$ can be excluded at the respective confidence level.}
\end{figure}
The sensitivity to any CP violation is shown in \figu{anycp} for Setup~I
(upper left-hand side plot), Setup~II (upper right-hand side plot), Setup~II 
(lower left-hand side plot), and the neutrino factory scenario (lower right-hand 
side plot) at the 1, 2, 3, 4, and 5 $\sigma$ confidence level from bright 
grey/yellow (1$\sigma$) to red/dark grey (5$\sigma$). Sensitivity to any CP 
violation is given for a pair of true values $\stheta$-$\deltacp$ if the CP 
conserving values $\deltacp=0$ and $\deltacp=\pi$ do not fit the simulated 
reference data if all correlations and degeneracies are taken into account. 
It is known, that the standard neutrino factory suffers from the sign($\Delta
m^2_{31}$)-degeneracy in some areas of the parameter space ($\stheta\approx 
10^{-2.5}$ and $\deltacp\approx -\pi/2$), because of the so-called 
``$\pi$-transit'',~\ie the degenerate solution fitted with wrong sign of $\ldm$ 
contains the CP conserving value for $\deltacp=\pi$ (see~\cite{Huber:2002mx} for 
details). As can be seen from \figu{anycp}, Setup~I suffers strongly from
correlations and degeneracies at larger true values of $\stheta$ whereas
Setup~II performs better. Within the interval $\deltacp \in [-\pi,0]$ Setup~II does not 
suffer from any correlations and degeneracies anymore and gives better results 
than the neutrino factory in the same interval. In the interval $\deltacp \in 
[0,\pi]$ Setup~II and the neutrino factory perform in a comparable manner, only 
for larger true values of $\stheta\gtrsim 10^{-2}$ the neutrino factory looses
sensitivity to CP violation for values of $\deltacp$ near the CP conserving values. 
This effect is due to the uncertainty of the matter density along the baseline 
which strongly affects the performance of a neutrino factory at large values 
of $\stheta$ because of the very long baseline. The best sensitivity to any CP 
violation is found for Setup~III. Here, the combination of data from the first 
and second oscillation maximum can resolve the degeneracies that appear at the 
baseline of L=600km for Setup~I. Additionally the sensitivity to CP violation of 
Setup~III reaches to significant smaller values of $\stheta$ at the maximally CP 
violating values $\deltacp=\pm \pi/2$. We checked that, as also was the case for 
sensitivity to $\stheta$, a combination of 2 years at $\gamma=900$ and 8 years at 
$\gamma=2500$ would also already allow to give this performance.   
\begin{figure}[t!]
\begin{center}
\includegraphics[width=9cm]{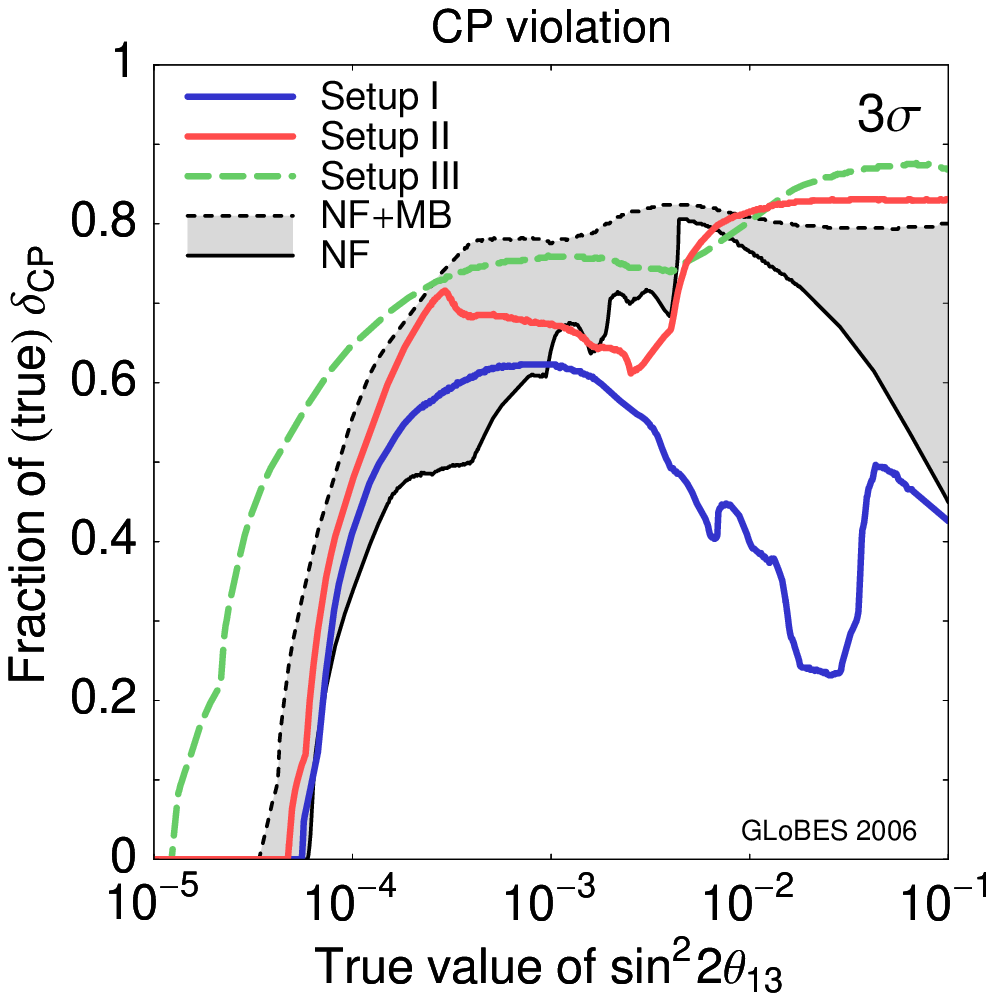}
\end{center}
\mycaption{\label{fig:fraction} Comparison of the fraction of (true)
$\deltacp$ for which CP violation can be established at the 3$\sigma$
confidence level as a function of the true value of $\stheta$ at the 
monobeam scenarios Setup~I, Setup~II, Setup~III. The solid black line is for a standard
neutrino factory while the dashed line is for an optimized neutrino factory with a second
detector at the magic baseline.}
\end{figure}
The results from \figu{anycp} are finally summarized in \figu{fraction}. The 
fraction of $\deltacp$ parameter space where sensitivity to any CP violation 
is given at the 3$\sigma$ confidence level is shown as a function of true 
$\stheta$ for the considered scenarios Setup~I, Setup~II, Setup~III and a neutrino factory.
The performance of the standard neutrino factory is indicated by the black solid line. However, we also show the performance of an
optimized neutrino factory scenario, where in addition to the
standard golden channel measurements at the baseline $L\sim3000$km a second 50kt Magnetized detector is installed approximately at
the magic baseline $L=7500$km. As can be seen in \figu{fraction}, the performance of the neutrino factory setup is significantly
improved. Note, that a CP
fraction of 1 can never be achieved, since values near the CP conserving
values can never be distinguished due to finite statistics.

\section{Summary and conclusions}

We have analyzed the potential of high gamma neutrino beams from electron 
capture decays of $ ^{110}_{50}\mathrm{Sn}$ isotopes directed towards a large 
Water Cherenkov detector with a fiducial mass of 500~kt. The resulting 
neutrino beam would be completely flavor pure and only consist of electron 
neutrinos. The achievable resolution in the energy reconstruction in such 
a scenario can be significantly more precise than from the usual energy 
reconstruction in Water Cherenkov detectors, since it is performed by the 
position measurement within the detector. The aim of this work was to 
estimate the potential and requirements of such scenarios to resolve 
the correlations and degeneracies in the sensitivity to $\stheta$ and the 
sensitivity to any CP violation, only with their power in energy resolution 
abilities. This has been compared to the performance at a neutrino factory, 
where the combination 
of neutrino- and anti-neutrino running is used to resolve correlations and 
degeneracies. We have compared three monobeam setups, two of them with a 
different energy window at different locations respective to the first 
oscillation maximum. Setup~I at a baseline of L=600km and $\gamma=2500$ has 
been chosen such, that the energy window of the analysis is directly 
located at the first oscillation maximum, but due to this choice the energy 
window is not broad enough to cover the whole oscillation maximum. Setup~II 
at a baseline of L=250km and $\gamma=2000$ on the other hand has a broader 
energy window which is located at higher energies as the oscillation maximum. 
In comparison to Setup~I this setup gains from the broader energy window and 
the larger statistics due to the smaller baseline. The baseline of L=600km 
also allows to take data at the second oscillation maximum, which is at this 
baseline already located at energies above the Cherenkov threshold of muons. 
Therefore Setup~III combines a measurement at the first oscillation maximum 
($\gamma=2500$ as in Setup~I) and the second oscillation maximum ($\gamma=900$), 
5 years data taking each. For the exposure of the setups it has been assumed
to have a running time of 10 years at a number of $10^{18}$ decays per year.
This number is hard to obtain because of the relative long lifetime of the 
$ ^{110}_{50}\mathrm{Sn}$ isotopes and an enhancement of the electron 
capture rate has to be achieved. However, concerning the 
sensitivity to $\stheta$ we found that this number is required to evolve 
capabilities to start resolving the correlations but still the pure superb 
energy resolution and the high statistics alone cannot compete with the 
sensitivity to $\stheta$ at a standard neutrino factory with a 50~kt MID 
detector at a baseline of L=3000km and a parent muon energy of $\mathrm{E_\mu=50GeV}$ 
because the degeneracies cannot completely be resolved. On the other hand the neutrino 
factory also suffers strongly from degeneracies and additional data from the 
silver channel, the magic baseline or lower energies (maybe with a hybrid 
detector) would be required. Setup~III on the other hand with the combination 
of data from first and second oscillation maximum performs well in
resolving the correlations and degeneracies. It gives a better sensitivity 
$\stheta\lesssim 2.5\cdot10^{-4}$ at the 3$\sigma$ confidence level. When it 
comes to the sensitivity to any CP violation the performance of the monobeam 
setups is more impressive than the performance concerning the sensitivity to 
$\stheta$. Setup~I still suffers significantly from the degeneracies while 
Setup~II reaches sensitivity in a quite large part of the parameter space and no
negative effects from degeneracies could be observed. Finally, Setup~III showed 
very good abilities to establish CP violation in a very large part of the parameter 
space and all degeneracies coming from the measurement in Setup~I can be resolved 
due to the data from the second oscillation maximum although no information from 
a anti-neutrino running is included. However, one has to note that the requirements 
to achieve such a performance,~\ie the very large acceleration factors of the 
isotopes, the high number of isotope decays per year, and the very low beam divergence of the stored isotopes of $\mathcal{O}(1\ \mu
\mathrm{rad})$ are extreme.

\subsection*{Acknowledgments}

M.~R. would like to thank Marc-Thomas Eisele for useful discussions.
M.~R. is supported by the Graduiertenkolleg 1054 of Deutsche Forschungsgemeinschaft. 
J.~S. is partially supported by Grant-in-Aid for Scientific Research on Priority Area
No.~1774013 and No.~18034001.

\newpage

\begin{appendix}

\section{Relativistic transformations}

\subsection{Neutrino energy}

In the following considerations, the primed quantities are defined 
as the quantities in the laboratory frame,~\ie the rest frame of 
the detector and the quantities without a prime represent those in 
the rest frame of the electron capture decays in which the neutrinos 
are produced.
\begin{center}
\begin{minipage}{\textwidth}
\includegraphics[height=3.3cm]{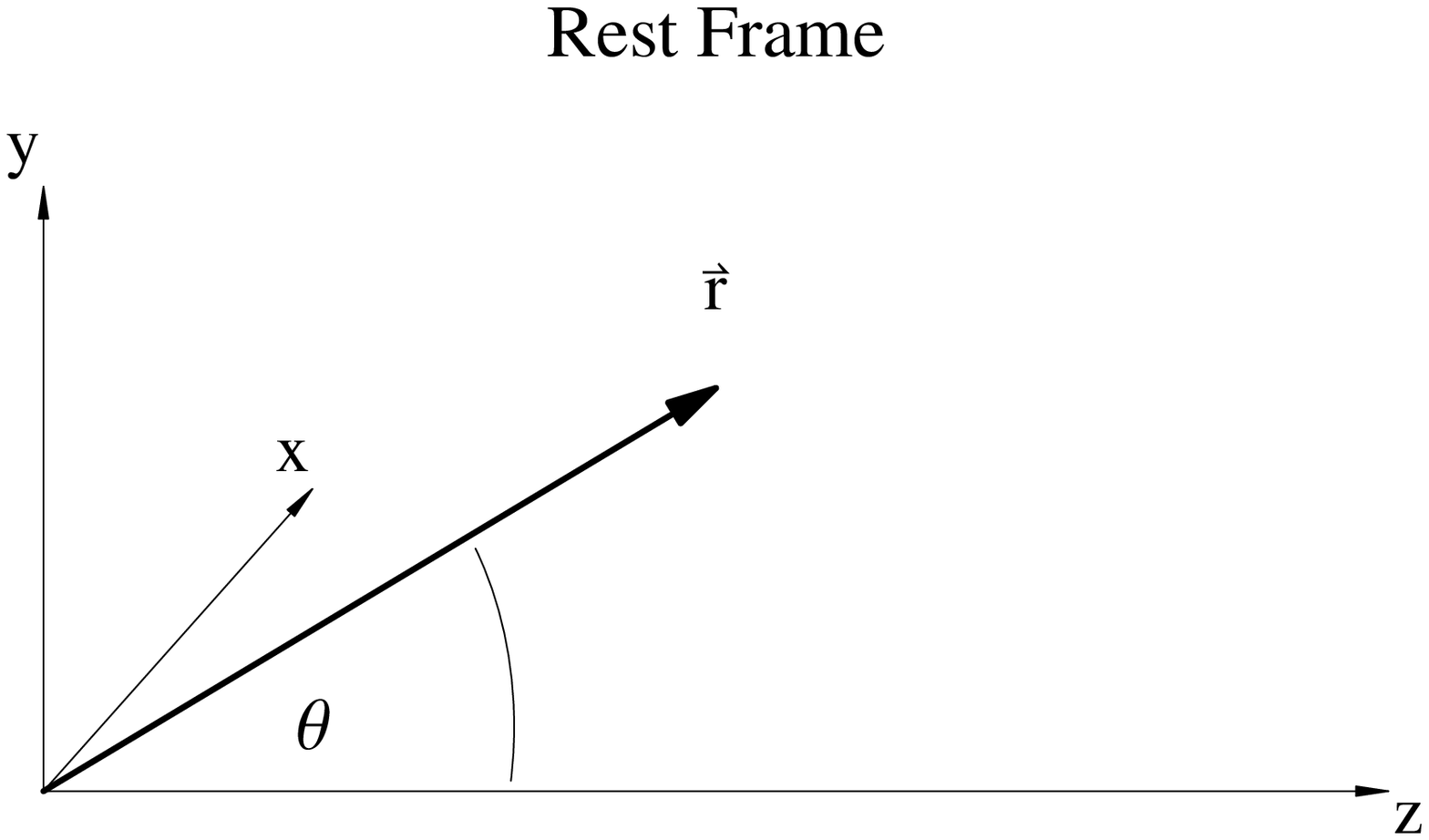} 
\includegraphics[height=3.3cm]{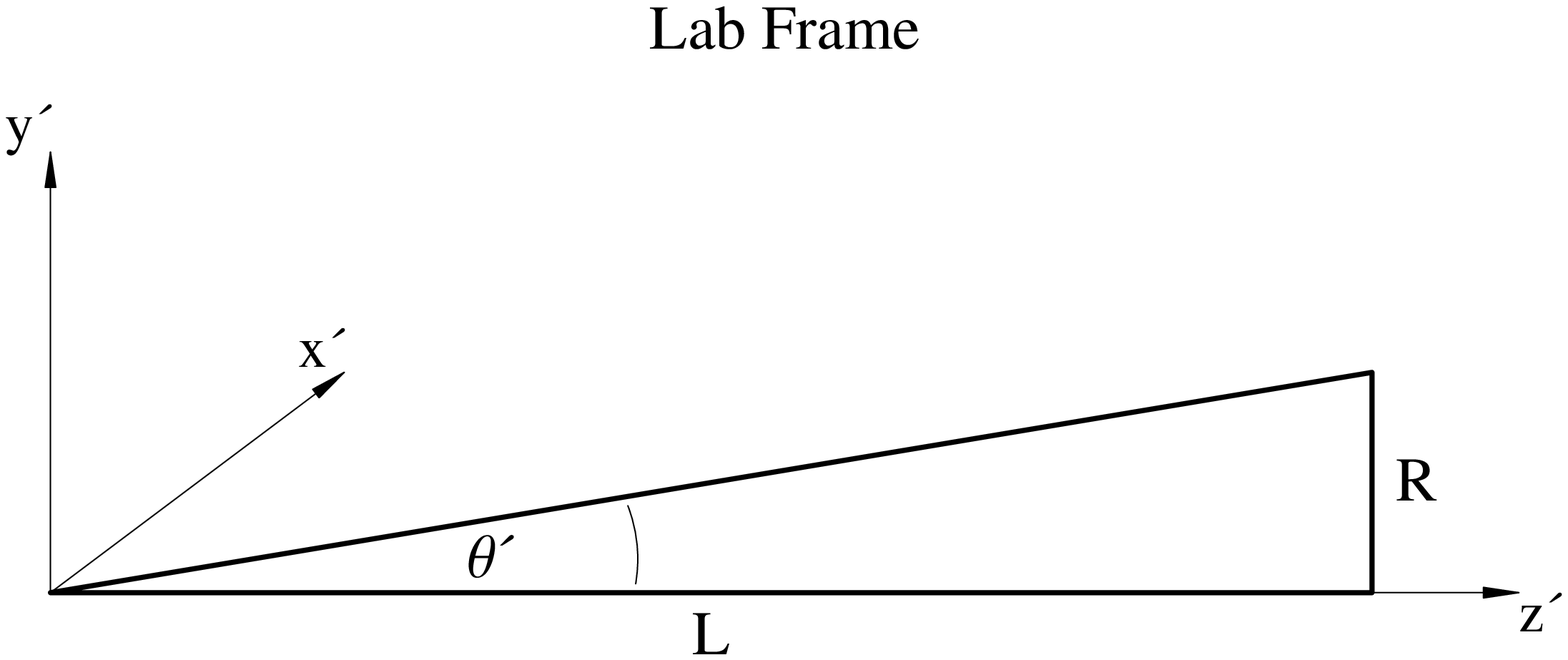} 
\end{minipage}
\end{center}

In the rest frame of the decays the neutrinos are produced at an
energy $Q$ and with an uniform angular distribution of momentum. 
Since for the considered mother nuclei $^{110}_{50}\mathrm{Sn}$ the endpoint 
energy is Q~=~267~keV ($m_\nu \ll Q$), the neutrino mass can be 
neglected:
\begin{eqnarray}
Q^2 = p^2 + m^2 \approx p^2.
\end{eqnarray} 
So, for a neutrino that is emitted in the direction $\vec{r}$ the
four-momentum in the rest frame of the decay is given by
\begin{eqnarray}
p^\mu = \left( \begin{array}{c}
Q \\
\\
Q \vec{e}_r \\
\\
\end{array}
\right) 
=
\left( \begin{array}{c}
Q \\
Q \; \sin\theta \; \cos\phi \\
Q \; \sin\theta \; \sin\phi \\
Q \; \cos\theta  
\end{array}
\right).
\end{eqnarray}
Since the problem is $\phi$-symmetric, we can choose $\phi = 0$ 
for the sake of simplicity and the four-momentum of the neutrino 
in the rest frame can be written as
\begin{eqnarray}
p^\mu = 
\left( \begin{array}{c}
Q \\
Q \; \sin\theta \\
0 \\
Q \; \cos\theta  \\
\end{array}
\right).
\end{eqnarray}
The boost is in the $z$-direction, and after the transformation 
the energy of the neutrino in the lab frame becomes
\begin{eqnarray}
E' = p'^0 = \gamma \; Q (1+ \beta \, \cos\theta).
\end{eqnarray}

\subsection{Transformation of angles}

Now we want to derive the energy of a neutrino that hits the detector 
at a baseline L and at the distance $R$ from the beam center,~\ie at 
an angle
\begin{eqnarray}
\cos\theta' = \frac{L}{\sqrt{L^2+R^2}} =
\frac{1}{\sqrt{\; 1+(L/R)^2}}.
\end{eqnarray}
The expression for the neutrino energy has to be found as a function
of the angle $\cos\theta'$, or respectively the radius $R$.

From $p^{\mu'}$ it is quite straight forward to find the transformation
of $\cos\theta$:
\begin{eqnarray}
\cos\theta'= \frac{\gamma Q (\beta+\cos\theta)}{\sqrt{(\gamma Q
(\beta+\cos\theta))^2+(Q\sin\theta)^2}}=\frac{\beta+\cos\theta}{1+\beta\cos\theta}
\end{eqnarray}
and in the other direction the transformation is given by
\begin{eqnarray}
\cos\theta=\frac{-\beta+\cos\theta'}{1-\beta\cos\theta'}.
\end{eqnarray}
The transformation of $\phi$ is trivial $\phi=\phi'$ and therefore we find that
\begin{eqnarray}
\frac{d\Omega}{d\Omega'}=\frac{d\cos\theta}{d\cos\theta'}
\label{equ:Omegaratio1}
\end{eqnarray} 
with
\begin{eqnarray}
\frac{d\cos\theta}{d\cos\theta'}=\left[\gamma^2(1-\beta\cos\theta')^2\right]^{-1}
\label{equ:Omegaratio2}
\end{eqnarray} 
and the corresponding
\begin{eqnarray}
\frac{d\cos\theta'}{d\cos\theta}=\left[\gamma^2(1+\beta\cos\theta)^2\right]^{-1}.
\end{eqnarray} 
Now, the exact formula for the neutrino energy in the lab frame as a 
function of the lab frame quantities is found to be:
\begin{eqnarray}
E'(\cos\theta')=\frac{Q}{\gamma}\frac{1}{1-\beta\cos\theta'}
\label{equ:EinOmegaprime}
\end{eqnarray}
and
\begin{eqnarray}
E'(R)=\frac{Q}{\gamma}\left[1-\frac{\beta}{\sqrt{1+(R/L)^2}} \right]^{-1}.
\end{eqnarray}

\section{Calculation of event rates}

The initial neutrino beam consists only of electron neutrinos. In the
detector the muon neutrinos from the appearance channel will be
detected. The neutrino energy is maximal at the beam center ($R=0$) with
$E_{\mathrm{max}}=2\gamma Q$ and decreases to the outer regions of the
detector. We introduce an equidistant binning in $R^2$ to have more
balanced event numbers in the different bins, than would be the case for
equidistant binning in $R$. In the simulations, we use $k=100$ bins, so that
the largest bin appears in the beam center with approximately 10~m radius 
and the smallest bin is found at the outer edge of the detector with a width 
of approximately 50~cm, whereas the position measurement resolution is assumed 
to be at least 30~cm, which is the vertex resolution estimated for 
fully-contained single ring events at \SK\cite{Ashie:2005ik} in the energy 
window of interest. In this work we have not introduced an additional smearing 
between the bins in the outer regions of the detector. However, if the 
vertex resolution cannot be optimized this binning turns out to be too 
narrow at the outer bins in the detector and smearing would have to be 
introduced to these bins or the width of the bins in the analysis 
would have to be customized. We checked, that going to an equidistant binning 
in $R^2$ with only 50~bins,~\ie bin sizes from 14~m to 1~m, or going to an 
equidistant binning in $R$ with 100~bins hardly changes the main results 
of this work.

For the usage within the \globes software, the radial binning is
translated to binning in energy, where the bins are not equidistant
anymore.

If $R_{\mathrm{max}}^2$ is divided in $k$ bins the edges of the bins are
\begin{eqnarray}
R_i^2 = R_{\mathrm{max}}^2-(i-1)\Delta R^2
\end{eqnarray}
with 
\begin{eqnarray}
\Delta R^2=\frac{R_{\mathrm{max}}^2}{k}.
\end{eqnarray}
Here always $R_i^2>R_{i+1}^2$ holds, so that the corresponding energy
bins are in the right ordering for \globes:
\begin{eqnarray}
E'(R_i^2)<E'(R_{i+1}^2).
\end{eqnarray}
Furthermore, within \globes for the calculations the mean value of each
energy bin is taken:
\begin{eqnarray}
E_i=\frac{E'(R_i^2)+E'(R_{i+1}^2)}{2}.
\end{eqnarray} 
Then, the appearance event number in one energy-bin is given by  
\begin{eqnarray}
N_i\simeq\epsilon_i \times P(L,E_i)_{\nu_e\rightarrow\nu_\mu}\times \frac{1}{L^2} 
 \frac{dn}{d\Omega'}(E'_i) \times \sigma(E'_i) \times N_{\mathrm{nuc,i}},
\label{equ:events}
\end{eqnarray} 
where $\epsilon_i$ is the signal efficiency in the corresponding bin, 
$P(L,E_i)_{\nu_e\rightarrow\nu_\mu}$ is the appearance
oscillation probability, $\frac{dn}{d\Omega'}(E'_i)$ is the angular neutrino flux,
$\sigma(E'_i)$ is the charged current cross section per nucleon, and 
$N_{\mathrm{nuc,i}}$ is the number of nucleons within the geometrical size of 
the i-th bin: 
\begin{eqnarray}
N_{\mathrm{nuc,i}}=\Gamma_i \times
\frac{M_{\mathrm{det}}}{m_{\mathrm{nuc}}}=\frac{1}{R_{\mathrm{max}}^2}
\left[ R_i^2-R_{i+1}^2\right]
\times \frac{M_{\mathrm{det}}}{m_{\mathrm{nuc}}}=\frac{1}{k} \times
\frac{M_{\mathrm{det}}}{m_{\mathrm{nuc}}}.
\label{equ:nucleons}
\end{eqnarray}
Here $\Gamma_i$ is the fraction of all number of nucleons that have to
be considered in the i-th energy bin.

Since the neutrino flux in the rest frame of the decays is uniformly
distributed, it can be written as
\begin{eqnarray}
\frac{dn}{d\Omega}=\frac{N_{\mathrm{decays}}}{4\pi},
\label{equ:DistributionInRest}
\end{eqnarray} 
where $N_{\mathrm{decays}}$ is just the number of decays,~\ie the number of
produced neutrinos.  The neutrino flux can now be found with 
\eqs~(\ref{equ:Omegaratio2}) and (\ref{equ:EinOmegaprime}):
\begin{eqnarray}
\frac{dn}{d\Omega'}_i=\frac{dn}{d\Omega}\frac{d\Omega}{d\Omega'}=
\frac{N_{\mathrm{decays}}}{4\pi}\left[
\gamma^2
(1-\beta\cos\theta'_i)^2\right]^{-1}=\frac{N_{\mathrm{decays}}}{4\pi}
\left(\frac{E'_i}{Q}\right)^2.
\end{eqnarray} 
Also, it is straight forward to show by using 
\eqs~(\ref{equ:Omegaratio1}), (\ref{equ:Omegaratio2}),
(\ref{equ:EinOmegaprime}), and (\ref{equ:DistributionInRest}) that 
\begin{eqnarray}
 dn&=&\frac{dn}{d\Omega}\frac{d\Omega}{d\Omega '}\frac{d\Omega '}{dE '}
dE' =\frac{N_{\mathrm{decays}}}{2\beta\gamma Q} dE',
\end{eqnarray}
\ie $\frac{dn }{dE'}$ is constant.

\end{appendix}

%
\bibliographystyle{./apsrev} \bibliography{./references}

\end{document}